\pdfoutput=1
\documentclass{llncs}
\usepackage{soul}
\usepackage{etoolbox}
\usepackage{hyperref}
\usepackage{amssymb}
\usepackage{graphicx}
\usepackage{amsmath}
\usepackage{booktabs}
\usepackage{multirow}
\usepackage{subcaption}
\usepackage{geometry} 
\usepackage[margin=1cm]{caption}
\begin{document}

\author{Charilaos Akasiadis\inst{1,}\textsuperscript{+} \and
Miguel Ponce-de-Leon\inst{2,}\textsuperscript{+} \and
Arnau Montagud\inst{2,}\textsuperscript{$\dagger$}  \and\\
Evangelos Michelioudakis\inst{1,3,}\textsuperscript{$\dagger$} \and
Alexia Atsidakou\inst{1,}\textsuperscript{$\dagger$}\and
Elias Alevizos\inst{1,}\textsuperscript{$\dagger$}\and\\
Alexander Artikis\inst{1,}\textsuperscript{$\ddagger$}\and 
Alfonso Valencia\inst{2,}\textsuperscript{$\ddagger$}\and
Georgios Paliouras\inst{1,}\textsuperscript{$\ddagger$}
}
\institute{Institute of Informatics and Telecommunications, NCSR `Demokritos', Agia Paraskevi, Greece \\{\scriptsize\email{\{cakasiadis,vagmcs,aatsidakou,alevizos.elias,a.artikis,paliourg\}@iit.demokritos.gr}}\\
 \and
Life Sciences Department, Barcelona Supercomputing Center, Barcelona, Spain\\
{\scriptsize\email{\{miguel.ponce,arnau.montagud,alfonso.valencia\}@bsc.es}}
\\
\and
Department of Informatics and Telecommunications, University of Athens, Athens, Greece\\
\textsuperscript{+,$\dagger$,$\ddagger$}Equal author contribution
}

\title{Parallel Model Exploration for Tumor Treatment Simulations}

\maketitle

\abstract{Computational systems and methods are often being used in biological research, including the understanding of cancer and the development of treatments. Simulations of tumor growth and its response to different drugs are of particular importance, but also challenging complexity. The main challenges are first to calibrate the simulators so as to reproduce real-world cases, and second, to search for specific values of the parameter space  concerning effective drug treatments.
In this work, we combine a multi-scale simulator for tumor cell growth and a Genetic Algorithm (GA) as a heuristic search method for finding good parameter configurations in reasonable time. 
The two modules are integrated into a single workflow that can be executed in parallel on high performance computing infrastructures.
In effect, the GA is used to calibrate the simulator, and then to explore different drug delivery schemes.
Among these schemes, we aim to find those that minimize tumor cell size and the probability of emergence of drug resistant cells in the future.
Experimental results illustrate the effectiveness and computational efficiency of the approach.
}

\section{Introduction}
Computational systems biology is a research field that combines mathematical and computational models, together with molecular data, in order to improve our understanding of biological systems~\cite{kohl_systems_2009}.
Among others, it has been applied to the study of cancer in an attempt to understand the behavior of tumors and predict treatment effectiveness~\cite{hornberg_cancer:_2006}.
In particular, computational simulations have been used to model the development and evolution of tumor cells, which is the result of many different interacting processes that occur at different scales across time and space. 
Take as an example the mutations and other DNA alterations that happen at the molecular level and can potentially damage the function of genes~\cite{vogelstein_cancer_2013}.
These mutated genes may have an impact on the correct functioning of cell processes, such as signal transduction and gene regulation. 
In turn, this can lead to the transformation of healthy cells into malignant or tumor cells~\cite{hanahan_hallmarks_2011}.
Therefore, modelling and simulating cancer is a challenging problem, due to the multi-scale nature of this complex multicellular disease~\cite{deisboeck_multiscale_2011}.
In parallel, realistic models are instrumental to the interpretation of biological experiments and the derivation of mechanistic explanations that can be translated into new experimentally testable hypotheses~\cite{an_closing_2010}.
This way, faster and zero-risk experiments can be performed in-silico to provide further insight to domain experts.

Focusing on drug treatment exploration, the systems biology community has developed various approaches for predicting novel targets and new drugs that can potentially stop or reduce tumor growth~\cite{du_cancer_2015}.
Each of these approaches may focus on a different cellular process associated with cancer, such as cell signalling~\cite{calzone_logical_2018}, or metabolism~\cite{yizhak_modeling_2015}.
However, tumor cells display a high degree of heterogeneity, which is strongly related to the capability of tumors to develop resistance to drugs, 
or to attacks from immune cells or changing environmental conditions~\cite{lipinski_cancer_2016,robertson-tessi_impact_2015}. 
For this reason, analytical model formulations that assume particular distributions and related parameters, or operate at a macroscopic level to derive closed formed solutions, are not always suitable~\cite{Khac}.
Instead, this has motivated the development of multi-scale agent-based approaches that integrate intracellular models into individual cell agents, allowing the simulation of heterogeneous populations and the modelling of how cell variability can affect a treatment outcome~\cite{metzcar_review_2019}.
Such agent-based models provide a balanced combination of features that suits well the modelling of biological tissues, as they combine: \textit{(i)} the ability to study the behavior of cell populations that have a direct correspondence to biological tissue phenotypes, \textit{(ii)} cell-level granularity that allows us to study mutations at the single-cell level, \textit{(iii)} intracellular models that can capture gene-level changes, and \textit{(iv)} explicit descriptions of the environment, which have a direct correspondence to laboratory experiments~\cite{montagud_systems_2021}.
Multi-scale models have been used to simulate the evolution of a tumor by taking into account the molecular details of individual cells and, thus, are valuable tools for performing in-silico experiments~\cite{ghaffarizadeh_physicell:_2018,letort_physiboss:_2018}.

Despite their power and versatility, multi-scale models have many different parameters (e.g., diffusion constants, rates of different cellular processes) whose values must be properly fine-tuned in order to reproduce biologically plausible simulations that are in line with real-world outcomes~\cite{babtie_how_2017}.
A subset of these parameters can be set, based on experimental measurements in the literature or in relevant databases. 
However, for most of them, there are no such real-world measurements and they need to be calibrated or inferred using indirect sources of information. 
A common practice is to employ optimization methods in order to find the parameter values that better explain the available experimental observations~\cite{tekin_simulation_2004}.
However, multi-scale models are complex and the optimization of model parameters on the data cannot be performed given the absence of a sufficient amount of data. 
As a result, the evaluation of a candidate set of parameter values must be performed by executing simulations that require significant time and computational resources~\cite{ozik_desktop_2016}.
The results of such simulations then need to be inspected visually by experts or compared against desired outcomes or gold standards, which are already known to be biologically plausible and interesting. Such expert knowledge is very scarce and corresponds usually to just a few specific cases and experimental configurations. 
Additionally, due to the size and complexity of the parameter space, an exhaustive search is prohibitively expensive.
Therefore, efficient methods for exploring such complex parameter spaces are required~\cite{ozik_high-throughput_2018}.

In this paper, we  re-designed and extended a multi-scale agent-based model to simulate the growth of a tumor spheroid~\cite{letort_physiboss:_2018}.
The framework allows to explicitly simulate the cell population without the need for any coarse-grained approximation. Our model takes into account the effects of the Tumor Necrosis Factor (TNF), a signalling molecule that binds to cell receptors and can trigger a wide range of different responses~\cite{li_structural_2013}.
In particular, TNF can induce death in cancer cells by activating specific downstream signalling pathways, thus restraining the growth of a tumor. The extended version of the multi-scale model includes an explicit submodel of the TNF-receptor dynamics. 
To simulate the effect of the drug, we use a Boolean network for modelling intracellular signalling and cell fate, which is a generic approach for cancer signalling that has already been validated experimentally. 
This model accounts for the emergence of resistance after long exposure periods and the mechanisms are explicit in the structure of the regulatory network.
The model incorporates experimental parameters, such as the doubling time and the volume of the specific cell line. However, the model also includes  unknown parameters that should be calibrated before using the framework to explore treatment strategies.
% (see fig. 1A). 
%
%\added{}
%\replaced{new}{old}
%\deleted{}

We have implemented a workflow on a High Performance Computing (HPC) cluster for the efficient calibration of  the model's unknown parameters, as well as for model exploration of treatment strategies that minimize tumor growth. 
In previous work, Letort et al.\cite{letort_physiboss:_2018} calibrated a first version of the model by integrating experimental data from Lee et al. (2016)~\cite{lee_nf-b_2016}. Herein, we use two simulated time trajectories reported by Letort et al. (2019) as ground truth to calibrate the values of  four unknown parameters introduced in the explicit TNF-receptor model. In other words, since our model has now an explicit submodel for the TNF-receptor dynamics, we used the simulation trajectories reported by Letort et al. as a gold standard to ``calibrate'' our extended version of the model. During the model calibration, we investigate different distance metrics for comparing the generated simulations against the gold standard, and we propose an informed search strategy, based on a Genetic Algorithm, which we compare against a plain Sweep search method.

Finally, we use the calibrated model and the implemented exploration workflow to investigate drug treatment strategies that minimize tumor size while avoiding the emergence of resistant cells in the long term, i.e., cells that become unresponsive to a drug or signal after being exposed to it for extended periods.\footnote{For a more in-depth study regarding this process, please refer to Calzone et al. 2010~\cite{calzone_mathematical_2010}.}
By analyzing our results, we can investigate the effect that drug dosage and frequency have on the overall cell survival. 
Furthermore, the parallelized framework that we propose can be easily re-used for simulating different experimental setups and cancer types, by simply replacing the Boolean model.\footnote{For instance, the systems biology community has reconstructed Boolean models of cell signalling networks for many different cancer types: \url{http://ginsim.org/models_repository}}
In summary, our contributions are:
\begin{itemize}
	\item The design and implementation of a multi-scale agent-based model for simulating tumor growth subject to TNF administration.
	\item The integration of the simulator into a parallel, HPC compatible model exploration workflow. The parallel workflow is used first for calibration, and then for searching effective drug treatment strategies.
	\item The incorporation of a distributed Genetic Algorithm into the workflow for searching the parameter space.% and its performance is compared to an uninformed Sweep search.
	\item Results show that our approach manages to find appropriate values for the parameters in question, resulting to effective treatment strategies that are able to reduce tumor growth, while avoiding the emergence of resistant cells.
	\item By parallelizing the workflow and employing the Genetic Algorithm, the overall complexity is reduced both in terms of time spent and computational resources required, compared to an uninformed Sweep search approach.
\end{itemize}

Therefore, the proposed approach reduces the computational requirements without compromising model quality. As a result, the cost of research on drug effects is reduced significantly. Altogether, our results show how multi-scale models, coupled with HPC-based model exploration workflows, can be used to perform large-scale \textit{in-silico} screening of treatment strategies, which can then be tested in \textit{in-vivo} models, accelerating the process of novel treatment discovery.

The remainder of this paper is structured as follows: Section 2 briefly discusses related work; Section 3 presents the model  that the multi-scale simulator is based on.
Then, Section 4 introduces the parameter exploration problem that we address, as well as the proposed distributed Genetic Algorithm approach.
Section 5 analyzes the experimental results, while Section 6 concludes and proposes directions for future work.

\section{Related Work}

Computational tools and methods are being applied extensively to solve problems of biology and chemistry.
A popular approach for the modelling of tissue response to therapies is to consider global optimization models and seek---or solve analytically---the parameter values that optimize particular objectives, such as the response of the tumor to a treatment.
For example, Lind et al.~\cite{lindetal} construct a probabilistic model and solve for the parameter values that maximize the probability that tumor control treatments will be complication-free. 
In another case, Lobato et al.~\cite{lobatoetal}, consider drug treatments as optimal control problems, which are solved by a multi-objective optimization differential evolution algorithm. 
The aim is to minimize the concentration of tumor cells, by using as few drugs as possible. 
Mohan et al.~\cite{mohanetal}, introduced a multi-step optimization procedure that takes into account various clinical factors and uses Simulated Annealing to search for better performing treatment plans. 
Zhang et al.~\cite{zhangetal} accounts for the drug properties and the changes induced in the tumor microenvironment, performing also in-vivo experiments to illustrate the effectiveness of the proposed treatment optimization strategy. 

In addition, modelling has been used to optimize tumor treatments that are not based on drugs.
Marino et al. \cite{marino2021openep} optimized the electroporation-based treatment of tumors, using a shared-memory implementation that takes advantage of parallel resources.
Ballo et al. \cite{ballo2019correlation} studied the optimal electric fields dosimetry in glioblastoma by creating magnetic resonance imaging models of patients, simulating the effect of different dose densities of electric fields and correlating them with the survival of patients.
In our work, we consider an agent-based model that simulates the growth of tumor spheroids. The basis of the proposed approach is a Boolean model of intracellular signals, which cannot be solved analytically to obtain the best drug treatments. 
To guarantee that simulations correspond to real cases, we fit the parameters of our model taking into account results from past trials. 
Then, we use the simulator to discover effective treatment strategies. 
For this purpose, we propose the use of a Genetic Algorithm, as that improves upon existing methods on the use of time and computational resources.

If we were to adopt a different approach for our problem, we would need to define a coarse-grained mean-field based on an ordinary differential equation that simulates the dynamics of the alive, apoptotic and necrotic cells. 
In the case of the alive cells, the equation should consider the effect of the TNF and it should also account for the emergence of resistance after prolonged periods of exposure. 
This would require introducing many assumptions on the different processes. 
Moreover, this would make the model very case-specific and hard to reuse in other contexts.
Instead, we implement our model using a general multi-scale modelling framework. 
The framework allows to explicitly simulate the population of cells without the need for any coarse-grained approximation. 
Furthermore, we use parameters measured for the experimental setup we are modelling, like the doubling time and the volume of the specific cell line use in the experiments. 
To simulate the effect of the drug, we use a Boolean network model for intracellular signalling and cell fate, which is a general model of cancer signalling that has already been validated experimentally. 
In our case, it accounts for the emergence of resistance after long periods of exposure and those mechanisms are explicit in the structure of the regulatory network. 
Since our model is implemented using a multi-scale framework, the different modelling components are modular and therefore other researchers can take advantage of it and adjust it to study other experimental models, just by changing parameter values on the configuration file. 
Therefore, besides the specific in-silico experiments performed and reported in this work, most components from our multi-scale model could be replaced or adapted to study other experimental systems in other conditions.

Genetic Algorithms have been used in a variety of cases~\cite{LO20181538,sliwoski2014computational}.
For instance, Sahlol et al.~\cite{sahlol2017evaluation} proposed the use of Neural Networks that had their weights optimized by a Genetic Algorithm to predict gene expression, related to the response to cisplatin-based chemotherapy.
Other researchers have applied Genetic Algorithms to find beneficial drug configurations for HIV patients, by also taking into account stochasticity in their model~\cite{8167912}. In that case, the cost function minimises the number of virus particles left in the organism.
In a very different domain, King et al.\cite{king2016genetic} employed a Genetic Algorithm to search very large mutant libraries for amino acid sequences that optimize certain desirable biochemical properties, such as the binding affinities for target cell receptors. The authors evaluated candidate sequences by running molecular dynamic simulations, a computational method which allows to estimate binding affinity in-silico.
Using their approach, the number of samples that need to be examined is significantly reduced and the quality of the acquired solutions improves.

Genetic Algorithms and Machine Learning approaches have also been extensively used to explore multi-scale models of multi-cellular systems, such as tumor growth. 
For example, Jagiella et al.\cite{jagiella_parallelization_2017} developed a parallel approximate Bayesian computation algorithm to parametrize multi-scale models of cells growing in a dynamically changing 3D nutrient environment. 
An early rejection mechanism is utilized in their work, i.e. a threshold on the objective function is set to determine if a non-interesting case appears and should be skipped, thus sparing computational resources.

Note that the case we are examining---i.e., searching a large parameter space---is challenging for many data-driven approaches, which require large amounts of annotated data, since such data is usually not available.
Instead, the approach proposed in this paper requires only some reference data (i.e., summarized results from just two past experiments) to calibrate a number of model parameters. The calibrated model can then be used to conduct simulations that correspond to in-silico experiments, where the outcomes may guide real world trials.

Recently, Ozik et al. have integrated the mechanistic 3D multicellular simulator PhysiCell with the model exploration platform EMEWS\cite{ozik_high-throughput_2018} and have used it to adaptively sample control parameters that maximize cancer regression in an agent-based model of immunosurveillance against heterogeneous tumors~\cite{Ozik2019}.
EMEWS (Extreme-scale Model Exploration With Swift) is a framework that enables the parallel execution of multiple model exploration tasks.
An Active Learning approach is used to explore the parameter space and discover optimal cancer regression regions for the parameters. The results of active learning were compared against those of a GA search in the parameter space. 
The Active Learning~\cite{Ozik2019} and Approximate Bayesian Computation~\cite{jagiella_parallelization_2017} may also be of use in our method as EMEWS components, but they come with a high computational cost.

In this paper, we develop a multi-scale agent-based model of a tumor spheroid using PhysiCell and provide the cell agents with an intra-cellular signal transduction model. 
The signalling model is used to compute cell responses to perturbations such as the presence of signalling molecules and drugs, which in our case is the Tumor Necrosis Factor (TNF) protein. 
We integrate this cancer model into the model exploration framework EMEWS in line with existing approaches in the literature~\cite{ozik_high-throughput_2018}. 
However, we develop a different workflow composed of two stages:
First, we calibrate the bio-physical parameters of our model. 
The calibration employs a genetic search for the optimal parameter values, guided by the set of bio-physical parameters that optimize the fitting to selected gold standards. 
To measure the distance between simulation time-series and gold standards, we test different distance metrics. 
In the second stage, we use the calibrated models to explore tumor reduction strategies, based on the periodic injection of a signal molecule, the TNF, that is able to induce tumor cell death. 
In this stage the model needs to capture complexities, such as the fact that cells exposed to the signal for long periods can develop resistance to the administered drug and evade death. 
Therefore, the proposed method for model exploration searches for TNF supply strategies that maximize tumor regression, while avoiding the emergence of cells resistant to the treatment.

\section{Simulation of the Biology Mechanics}
\label{sec:approach}
In this section, we describe the multi-scale model of 3D tumor spheroids, which is used to study and optimize treatment strategies that reduce the tumor size. Herein, a treatment corresponds to administering an amount of tumor necrosis factor (TNF) into the simulated microenvironment. In general, the signal triggered by the binding of the TNF to the cell's receptor will induce death in cancer cells, through either of two alternative mechanisms: Necrosis (NonACD) or Apoptosis. Nevertheless, after prolonged periods of exposure to the stimulus, cells find a way to bypass the death-inducing signal of the TNF and become resistant to the effect of the molecule~\cite{lee_nf-b_2016}. As a consequence, treatment strategies based on shorter ``pulses'' of TNF have been proposed to avoid the emergence of resistant cells (for further details, we refer the reader to the work of Calzone et al. (2010)~\cite{calzone_mathematical_2010}). In the following we present the details of our model's implementation.

We implemented our multi-scale model using the PhysiCell framework~\cite{ghaffarizadeh_physicell:_2018} with PhysiBoSSv2.0~\cite{letort_physiboss:_2018}.\footnote{The PhysiBoSSv2.0 is available at: \url{https://github.com/PhysiBoSS/PhysiBoSS}} PhysiCell is an open-source physics-based cell simulator for 3D multicellular systems that allows us to study many interacting cells in dynamic tissue microenvironments~\cite{ghaffarizadeh_physicell:_2018}. Cell mechanics are simulated using mechanical equations with default parameters from PhysiCell. The microenvironment is modelled using BioFVM~\cite{ghaffarizadeh_biofvm:_2016}, a solver for partial differential equations that can efficiently simulate key cell processes, such as secretion, diffusion, uptake, and the decay of multiple substrates in large 3D domains. PhysiBoSSv2.0 is an extension of PhysiCell that enhances the modelling capabilities by allowing simulations of Boolean models of regulatory networks within each individual cell-agent.
The computational complexity of the simulations scales linearly with respect to the number of cells ($\mathcal{O}(n)$).

\begin{figure*}
\centering
    \includegraphics[width=0.7\textwidth]{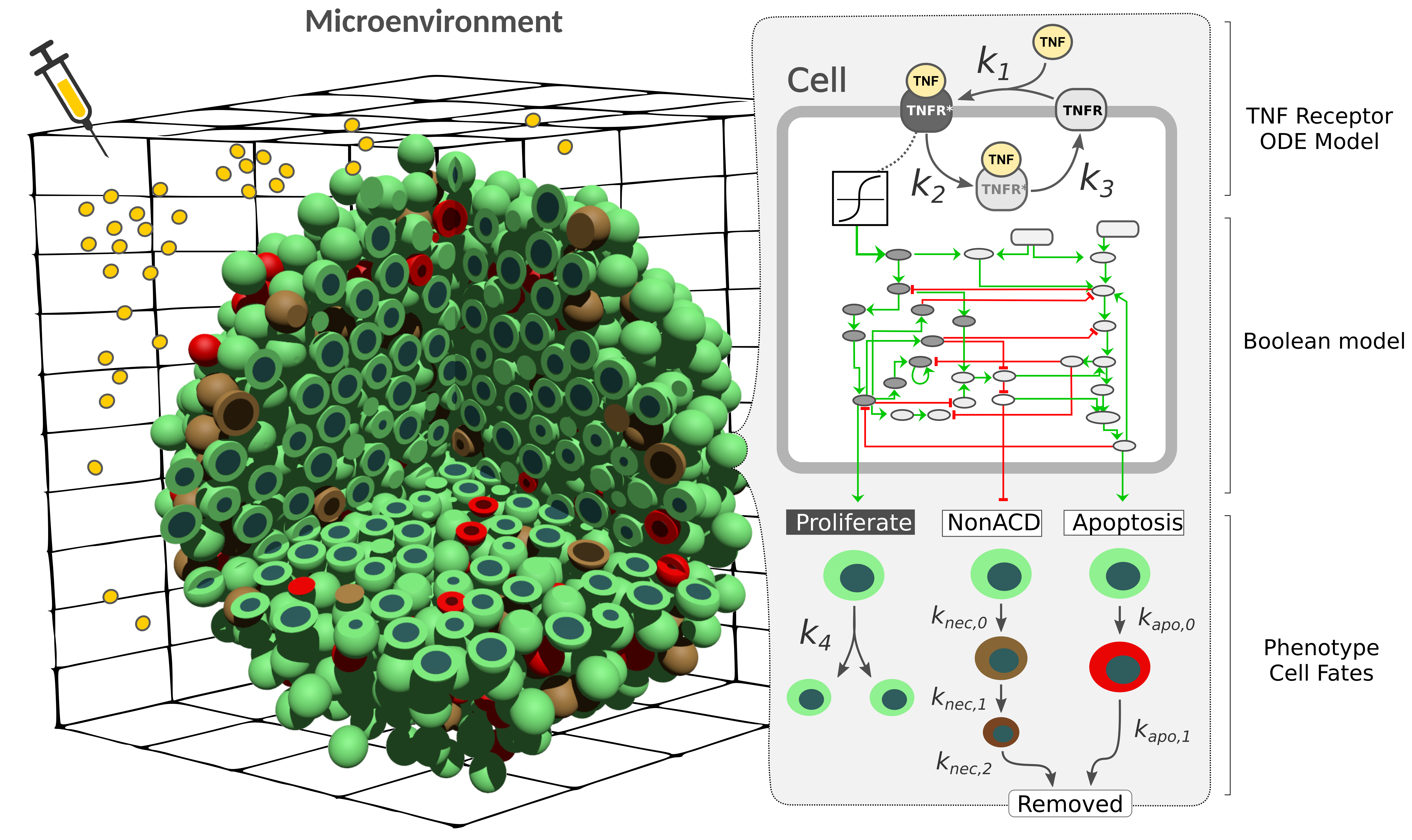} 
    \caption{On the left side, the figure represents an agent-based population of cells growing in a defined microenvironment. On the right side, the figure shows the intracellular models that rule the behavior of the cell agents. The models include: the TNF receptor model; the cell regulatory Boolean model; a transfer function to connect the two; the cell cycle model; and the death models for apoptosis and necrosis.}
    \label{fig:fig_model}
\end{figure*}

Figure~\ref{fig:fig_model} shows a schematic representation of our model that represents a tumor spheroid composed of a population of cells proliferating in a defined microenvironment. 
The microenvironment is modelled as a 3D domain which includes two diffusive molecules (or densities), one corresponding to oxygen (required for cell proliferation) and another one corresponding to the tumor necrosis factor (TNF). The diffusion, secretion, and import of the different densities are simulated using PhysiCell's standard partial differential equation solver, the BioFVM~\cite{ghaffarizadeh_biofvm:_2016}. In the case of  oxygen, we used the PhysiCell default configuration which corresponds to the oxygen level in normal conditions. On the other hand, the TNF supply depends on the treatment strategy used to counter tumor growth and thus, is a parameter subject  to optimization.

At the cell level, each individual tumor cell is modelled as an individual agent that has different internal sub-models, which represent and simulate the known molecular mechanisms that rule its behavior. As depicted on the right side of Fig.~\ref{fig:fig_model}, we can distinguish three intracelluar sub-models, one for modelling the TNF-receptor dynamics, one for modelling the cell regulatory network and one for the phenotype cell fate models. The TNF Receptor (TNFR) dynamics, i.e. the way TNF molecules attach to cells, is modelled using a set of differential equations that account for the experimental characterization of the TNFR binding dynamics described in in-vivo cells~\cite{fischer_ligand_2011} and considers \textit{(a)} the binding of the TNF to the cell receptor ($k_{1}$), \textit{(b)} the internalization of the TNF-receptor complex ($k_{2}$), \textit{(c)} the recycling of the receptor ($k_{3}$), and \textit{(d)} the cell growth rate ($k_{4}$, not shown in the figure). Cell receptors are proteins with the function of sensing different stimuli and transmitting signals by activating downstream regulatory pathways that end up affecting the cell fate. In our model, the TNF is the signalling molecule that binds to TNFR (receptor) and can trigger downstream effects. The binding process has an associated rate or kinetic constant ($k_{1}$).  After binding and triggering its effect, the complex formed by the TNF bound to the receptor gets internalized into the cell (endocytosis) at a given rate ($k_{2}$), and once inside the cell, the TNF is degraded and the receptor recycled for its further reuse ($k_{3}$). The equations below describe the sub-model:
\begin{equation}
  \begin{aligned}
  & \frac{[R_e]}{dt} =  - k_{1} [R_e][TNF] + k_{3} [R^{*}_{i}] \\
  & \frac{[R^{*}_{e}]}{dt} = k_{1} [R_{i}][TNF] -  k_{2} [R^{*}_{e}]  \\
  & \frac{[R^{*}_{i}]}{dt} = k_{2} [R^{*}_{e}] - k_{3} [R^{*}_{i}] \\
  \end{aligned}
\end{equation}

In the above equations $[R]$, $[TNF]$ and $[R^{*}]$ are the concentrations of the receptor, the TNF and the complex, respectively. The subscripts $i$ and $e$ denote internalized and external TNF receptor complexes. Within each cell, the TNF receptor model is simulated numerically using the backward Euler method and the numerical integration is conducted at the same time scale used to solve the diffusion. 

Then, when the TNFR complex concentrations ($[R^{*}_{e}]$) reach a defined threshold, specific regulatory pathways are activated~\cite{fischer_ligand_2011}.  The signal propagation through the pathways is modelled using the Cancer Cell Fate Boolean model, a Boolean network model tailored to cancer regulatory networks~\cite{calzone_mathematical_2010}. This model accounts for the most relevant regulatory pathways in the cancer cells and for three different cell fates or phenotypes, named Proliferation (`Alive' cells), Apoptosis (programmed cell death, or `Apoptotic' cells), and Necrosis (non-apoptotic cell death, or `Necrotic' cells). The Boolean model cell is simulated stochastically using the MaBoSS method~\cite{Maboss_2012}. After updating the internal state of the Boolean network, the Cell Fate node values are used to further update the Phenotype Cell Fate. Finally, to model Phenotype Cell Fates, we used the standard ``live cells'' cycle model~\cite{Friedman090456}, where live cells proliferate with a variable birthrate ($k_4$), or enter into necrosis or apoptosis with given rates $k_{nec,0}$ and $k_{apo,0}$, respectively (see Phenotypes Cell Fates at the right side of Figure~\ref{fig:fig_model}). The equations describing the process of Proliferation, Necrosis (NonACD), and Apoptosis are the following:
\begin{equation}
  \begin{aligned}
  Prob(divide) &\thickapprox  k_{4}  \Delta t \\
  Prob(necrosis) &\thickapprox
  \begin{cases} 
      k_{nec,0}  \Delta t & n_{NonACD} = 0 \\
      1 & n_{NonACD} = 1
   \end{cases}
   \\
  Prob(apoptosis) &\thickapprox
  \begin{cases} 
      k_{apo,0}  \Delta t & n_{Apoptosis} = 0 \\
      1 & n_{Apoptosis} = 1
   \end{cases}
  \end{aligned}
\end{equation}
where $Prob(divide)$ is the probability of a cell starting division at a given time step, and $Prob(necrosis)$ and $Prob(apoptosis)$ are the probabilities of a cell starting necrosis or apoptosis at a given time step, respectively. Cell growth and death are modelled using the standard PhysiCell models and default parameters, with the exception of cell growth rate $k_{4}$ which is calibrated by the method proposed in this paper. Nonetheless, at a given time, any cell can start apoptosis or necrosis if the corresponding Boolean node ($n_{Apoptosis}$ or $n_{NonACD}$) is active in the Boolean model.

Having developed the simulator, we select sets of parameter values, both for the rates of the agent model (denoted as $k_1$, $k_2$, $k_3$, $k_4$), and the drug treatment characteristics that are used in the exploration step. The different treatment strategies are defined as TNF pulses of a given concentration, duration and frequency, corresponding to three numerical values that are subject to exploration ($k_5$, $k_6$, and $k_7$ in Table~\ref{tab:ranges}). Subsequently, we use an optimization via simulation approach to first obtain values for the four $k_i$ rates, and then to identify treatment strategies that minimize the number of alive cells and avoid the emergence of resistance, i.e. cells that have activated their survival mechanisms upon TNF reception and are deemed unresponsive to the TNF treatment.

\section{Parameter Exploration: Calibration and Discovery of Treatments}
\begin{figure*}
\centering
  \includegraphics[width=0.7\textwidth]{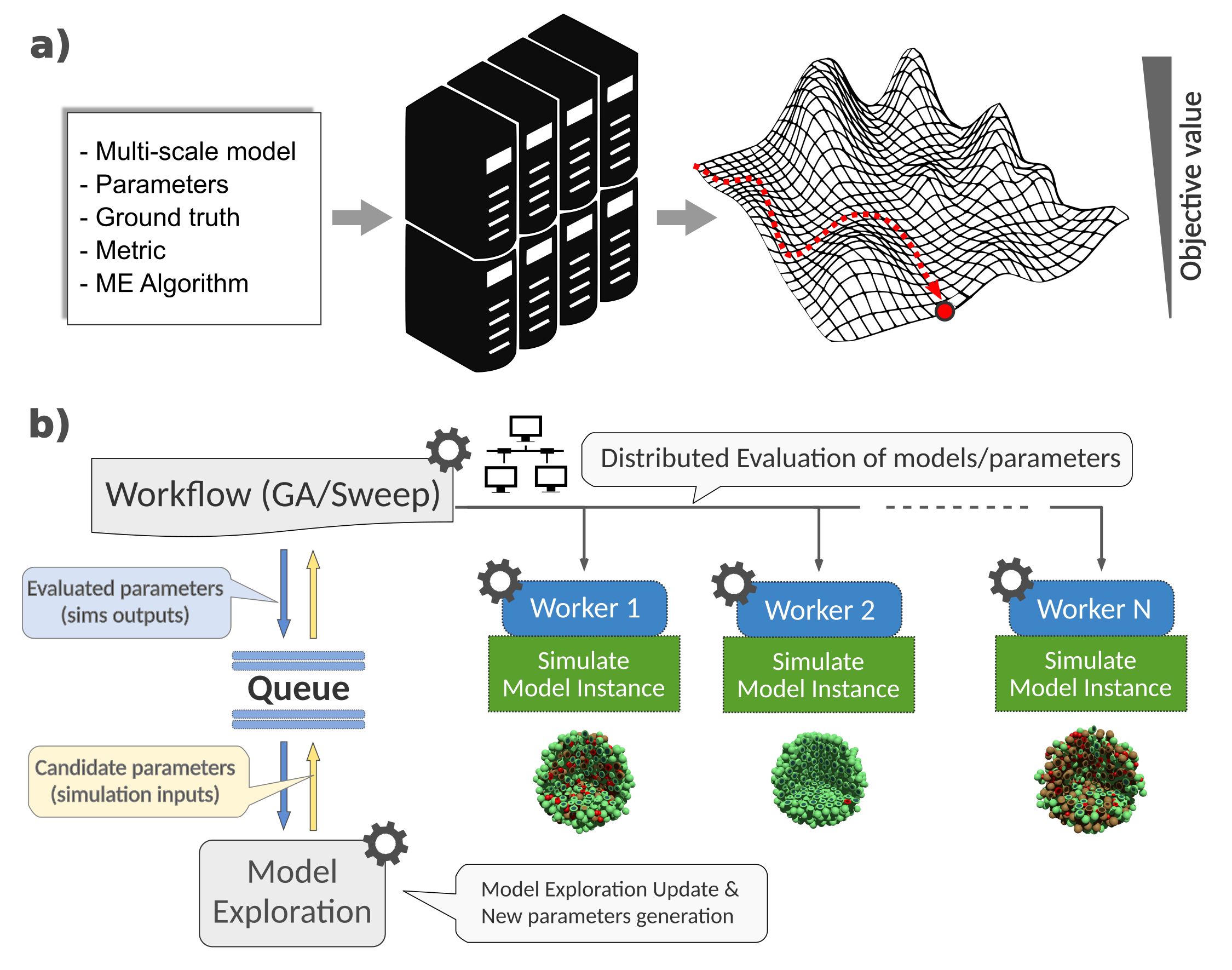} 
\caption{Overview of our model exploration approach. a) The multi-scale model is combined with Model Exploration (ME) in a High Performance Computing cluster
  to calibrate parameters and to explore treatment strategies. 
  b) Schematic overview of our EMEWS workflow, where the orchestrator (implemented in Swift/T) guides the execution of parallel simulations (PhysiBoSSv2.0 instances), aggregates evaluation results and communicates the results to the ME resident process (Sweep or Genetic Algorithm Search, implemented in Python) which learns from previous results and generates a new set of candidate parameters for their evaluation.}
\label{fig:fig_method}
\end{figure*}

Having described the simulator, we now present the parameter exploration framework that we propose.
Simulations are executed in parallel on an HPC infrastructure using the EMEWS framework~\cite{ozik_desktop_2016}. Once the results are obtained, a post-analysis is performed where only the relevant parts are examined, in our case the time-series of alive, apoptotic and necrotic tumor cells. 
Our implementation combines EMEWS for parallelizing the candidate solution evaluations and the DEAP Python library\footnote{\url{https://github.com/DEAP/deap}} for the Genetic Algorithm. 
The proposed approach follows the main principles of model exploration searching for parameter values that minimize desired metrics defined over the simulation results. 
A schematic overview of the approach is shown in Figure~\ref{fig:fig_method}.

A standard way to search for optimal parameter values could be to use a Sweep search method.
However, given the large size of the search space, a more efficient solution is needed, such as a Genetic Algorithm~\cite{russell2016artificial}.
The latter can search the space in a more informed manner, based on a fitness function that evaluates candidate solutions.
This way, we expect to achieve better results, using significantly fewer computational resources when compared to the exhaustive Sweep search.

\subsection{Configuring the multi-scale model of tumor growth}
To properly calibrate the simulator, it is crucial to determine the values for \textit{(i)} the TNFR binding rate, \textit{(ii)} the TNFR endocytosis rate, \textit{(iii)} the TNFR recycling rate, and \textit{(iv)} the cell growth rate. 
To ease notation, we refer to these parameters as $k_i$, $i\in \{1,2,3,4\}$, respectively.
Once the simulator has been calibrated, we can then explore the effectiveness of various drug treatment policies.
In this second step, the $k_i$ values correspond to the respective drug treatment configuration parameters, i.e. \textit{(i)} the TNF Administration Frequency, which shows how often the drug is injected and is measured in minute intervals, \textit{(ii)} the TNF Duration, dictating for how long the drug is administered, and \textit{(iii)} the TNF Concentration. These are denoted as $k_i$, $i\in \{5,6,7\}$, respectively.
The allowed value ranges for these variables are shown in Table~\ref{tab:ranges}.
\begin{table}
\centering
\caption{Ranges of the PhysiBoSS parameters. The first 4 are fitted during the calibration step, while the other 3 are explored during the drug discovery process. These ranges are estimated empirically by domain experts with long experience in this field of research.}
\label{tab:ranges}
\begin{tabular}{|l|c|c|}
\hline
Parameter&Min&Max\\
\hline
$k_1$: TNFR Binding rate& 0.01 & 1\\
$k_2$: TNFR Endocytosis rate& 0.01 & 1\\
$k_3$: TNFR Recycling rate& 0.01 &1\\
$k_4$: Cell growth rate&0.0001&0.002\\
$k_5$: TNF Administration Frequency&10&1200\\
$k_6$: TNF Duration&5&30\\
$k_7$: TNF Concentration&0.005&0.4\\
\hline
\end{tabular}
\end{table}

\subsection{Fitness function} \label{subsection:obj}
The suitability of specific configurations for the parameters $k_i$, $i\in \{1,2,3,4\}$ can be assessed by comparing the resulting simulations against some gold standard simulations that are considered ground truth. 
These gold standards constitute the time series of `Alive', `Apoptotic', and `Necrotic' cells, when applying TNF injections every 150 (TNF=150) and 600 (TNF=600) minutes.
In both injection strategies the duration is 10 minutes and the concentration of TNF is 0.02 TNF/$\mu m^3$.
Regarding the gold standard cases used in the optimization process, these may originate either from real world studies (possibly small-scale) or from other simulators that have been validated and shown to produce realistic results in the past.
Here, we use a stable version of the PhysiBoSS simulator to generate gold cases.

The first goal of the optimization process is to determine the $k_i$, $i\in \{1,2,3,4\}$ values that produce simulations as close as possible to the ground truth set by the gold cases.
Once the model has been calibrated, the second set of $k_i$, $i\in \{5,6,7\}$ parameters are explored, to find good drug treatment policies that can benefit real-world trials. 
To address each of the two optimization cases, the following minimization objectives are employed: \textit{(i)} the sum of distances between the gold standard time series and the simulations produced by the set of $k_i$ values under examination, or \textit{(ii)} the number of `Alive' cells at the end of each simulation, indicating the effectiveness of treatment policies.

The main idea behind the distance-based fitness score is that, when a simulation matches the gold standard, the distance will be minimized. When the distance is large, the fitness score of the candidate solution will decrease its probability of being selected in the evolution process of the GA. In summary, the calibration step is used to fit the parameters, i.e to find the parameter values that minimize the ``distance'' between the results obtained by the simulations and the gold standards.

Formally, for a simulation of length $T$ time points, let $L_t^j$, $N_t^j$, and $P_t^j$, $t\in 1, ..., T$ be the normalized number of `Alive', `Necrotic', and `Apoptotic' cells, for each of $j\in \{$TNF=150,TNF=600$\}$  gold standard cases, and $l_t^j$, $n_t^j$, and $p_t^j$, $t\in 1, ..., T$ be the normalized numbers of corresponding cell categories from the simulations produced by certain $k_i$ values. 
The normalization of cell numbers is made by dividing each value of the time series (`Alive', `Apoptotic', and `Necrotic' cell counts) by the maximum number of `Alive' cells from the respective $j$ case. 
This has shown to be necessary in order to remove bias, since, in the typical case, the absolute cell counts of TNF=600 are larger, resulting to increased absolute distance.

In the calibration step, the value of the fitness function is given by:
\begin{equation}
	F_{cal}=\sum_{j}{||\mathbf{L}^j-\mathbf{l}^j||_D+||\mathbf{N}^j-\mathbf{n}^j||_D+||\mathbf{P}^j-\mathbf{p}^j||_D}
	\label{eq:fit}
\end{equation}
where $||.||_D$ denotes one of the three different distance functions tested. 
Equation~\ref{eq:fit} is thus the actual fitness function that the GA seeks to minimize.
In the first step of the calibration, we test the following distance metrics:
\begin{enumerate}
	\item \textit{Euclidean} distance.
	\item The phase distance, computed according to the \textit{Dynamic Time Warping} algorithm\cite{berndt1994using}.
	\item \textit{$L_1$-norm}, i.e. the absolute differences.
\end{enumerate}

For the drug discovery step where $k_i$, $i \in {5,6,7}$ are optimized, the fitness score is simply the number of alive tumor cells at the end of the simulation (non-normalized):
\begin{equation}
	F_{disc}=l^T
	\label{eq:fit2}
\end{equation}
In this step, the gold standards are not used, since we are already equipped with meaningful $k_i$, $i\in \{1,2,3,4\}$ values.

\subsection{Genetic Algorithm}
\begin{figure}
	\centering
    \includegraphics[width=\textwidth]{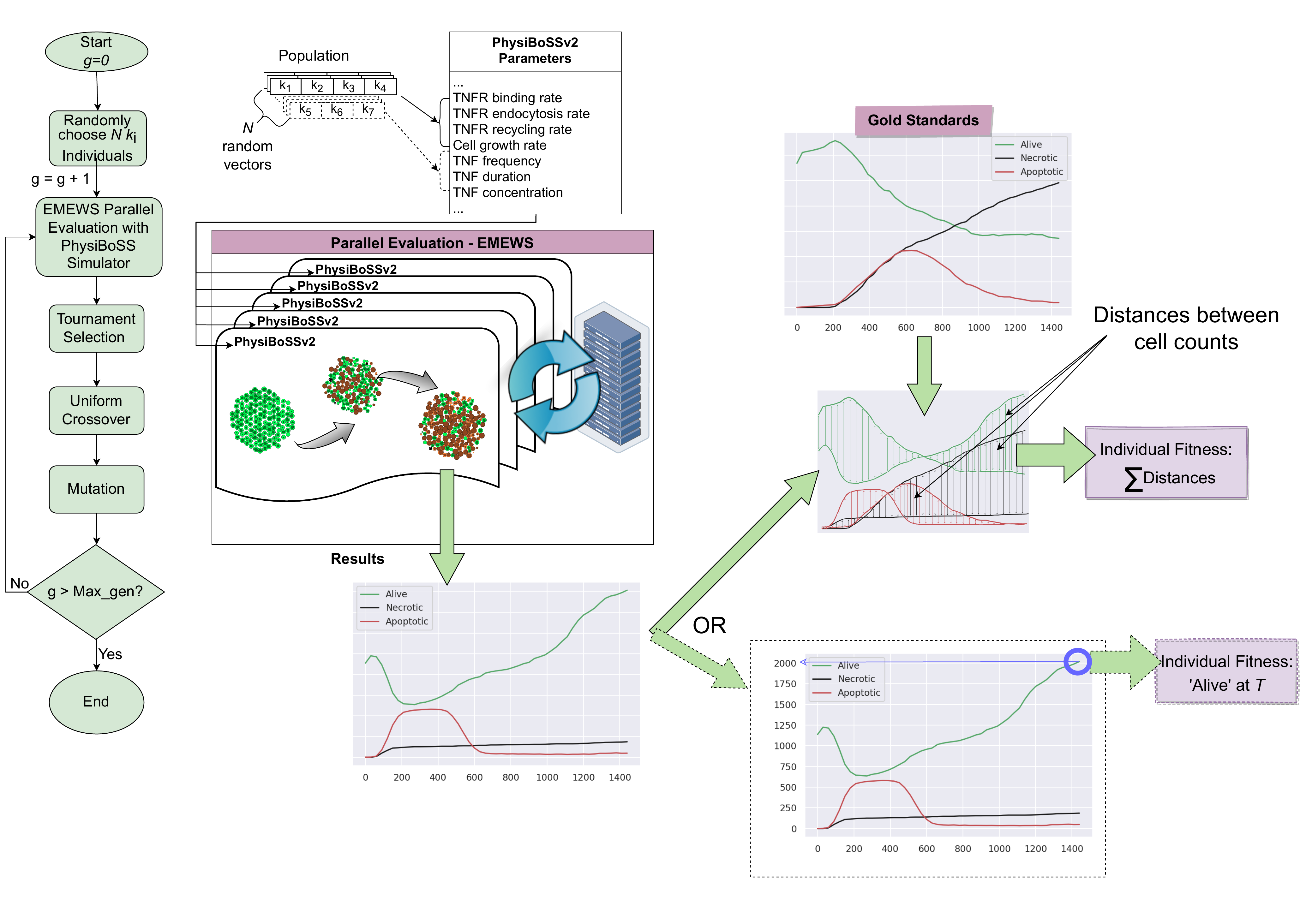}
    \caption{Genetic Algorithm flowchart. $g$ denotes generation index; $T$ denotes last simulation interval.}
    \label{fig:genetic_flow}
\end{figure}

The goal of the genetic search approach adopted by our model optimization method is to converge to optimal areas of the $k_i$ parameter
space and reveal high quality combinations of the desired parameters. 
A flowchart illustrating the execution stages of the proposed GA is shown in Fig~\ref{fig:genetic_flow}.

First, a random initial population of a given size is generated, i.e. random selections of $k_i$ values, which from now on will be referred to as individuals.
Individuals are represented as tuples of real values.
Depending on the optimization scenario (either calibration, or drug treatment exploration marked by a dashed border in Figure~\ref{fig:genetic_flow}), the values of some tuples are fixed. At the calibration stage we fix TNF time, duration, and concentration as dictated by the gold standard cases, while in the exploration phase we use the best binding, endocytosis, recycling, and growth rates discovered in the calibration trials. 
Given a particular parameter configuration, i.e., an individual, PhysisBoSSv2.0 performs the simulations and the results are evaluated according to the corresponding fitness function, either eq.~\eqref{eq:fit}, or eq.~\eqref{eq:fit2}.
Due to inherent simulator stochasticity, a feature that leads to slightly different cell count values in each run using the same $k_i$ values, simulations are repeated for the same individual.
The fitness score of each individual is the average of the scores produced by the duplicate runs.

The fitness of each individual is used to select individuals for the next generation of the GA.
Once the surviving individuals are selected, the crossover operator is applied on them, to generate new ones hence exploring alternatives that may lead to better fitness values. 
For similar purposes of exploration, mutation is randomly performed on each of the individuals, switching the value of a single parameter according to  a given probability.

This procedure, i.e. evaluation, selection, crossover, and mutation is performed repeatedly for a designated number of iterations \emph{Max\_gen}, or until other termination criteria are met.
In our implementation, the GA is configured to work with four different termination criteria that include reaching a maximum generation number, achieving a minimum fitness score, and converging to low population fitness variance or average scores for five consecutive generations. 
Finally, the result of the process includes the $k_i$ values that achieved the minimum fitness score, the score value itself, and the results of the corresponding PhysiBoSSv2.0 simulations during the evaluation for that particular individual. 
The latter comes in the form of time-series depicting the cell count of each cell category of interest. 
The main computational bottleneck in the presented GA workflow is the evaluation of each individual with PhysiBoSSv2.0.
For this reason, in the EMEWS workflow, each simulation is assigned to a number of threads, so that each part is effectively delegated to different processors.
This enables parallel execution, thus reducing the individual CPU computational burden.
Further computational savings are achieved by the proposed method in two ways: {\em (a)} by reducing the required simulations in both stages, and {\em (b)} by reducing the simulation time in the second stage, due to the use of more effective drug treatments.
Simulation time is shorter for effective treatments since fewer tumor cells survive.

 \subsection{Computational Complexity}
We are now looking closer at the computational savings introduced by the proposed framework in computational complexity terms.
As discussed in Section~\ref{sec:approach}, the complexity of a PhysiBoSS simulation increases linearly with the number of cells $n$ that take part in the analysis. In order to execute $m$ simulations sequentially, the time complexity would be $\mathcal{O}(m\cdot n)$. However, by taking advantage of the HPC infrastructure, and using $c$ processors, the complexity reduces to the order of $\mathcal{O}((m/c)\cdot n)$. This is the case for an exhaustive approach, like Sweep search. The complexity of the GA on the other hand, depends on its configuration, in particular on the population size $p$ and the number of generations $g$. Using $c$ processors for parallel execution, the time complexity of the GA is of order $\mathcal{O}(((p\cdot g)/c)\cdot n)$.
Therefore, in order for the GA to lead to computational savings compared to Sweep search, $(p\cdot g)$ should be smaller than $m$.  The experiments in Section~\ref{sec:exps} show that this is indeed the case. Moreover, it is worth noting that the parallelization of the GA is limited by $p$, i.e. the size of the population, as two individuals (simulations) in different populations cannot run in parallel.

\section{Experimental Results}
\label{sec:exps}
\begin{figure}[b]
\centering
   \includegraphics[width=0.5\textwidth]{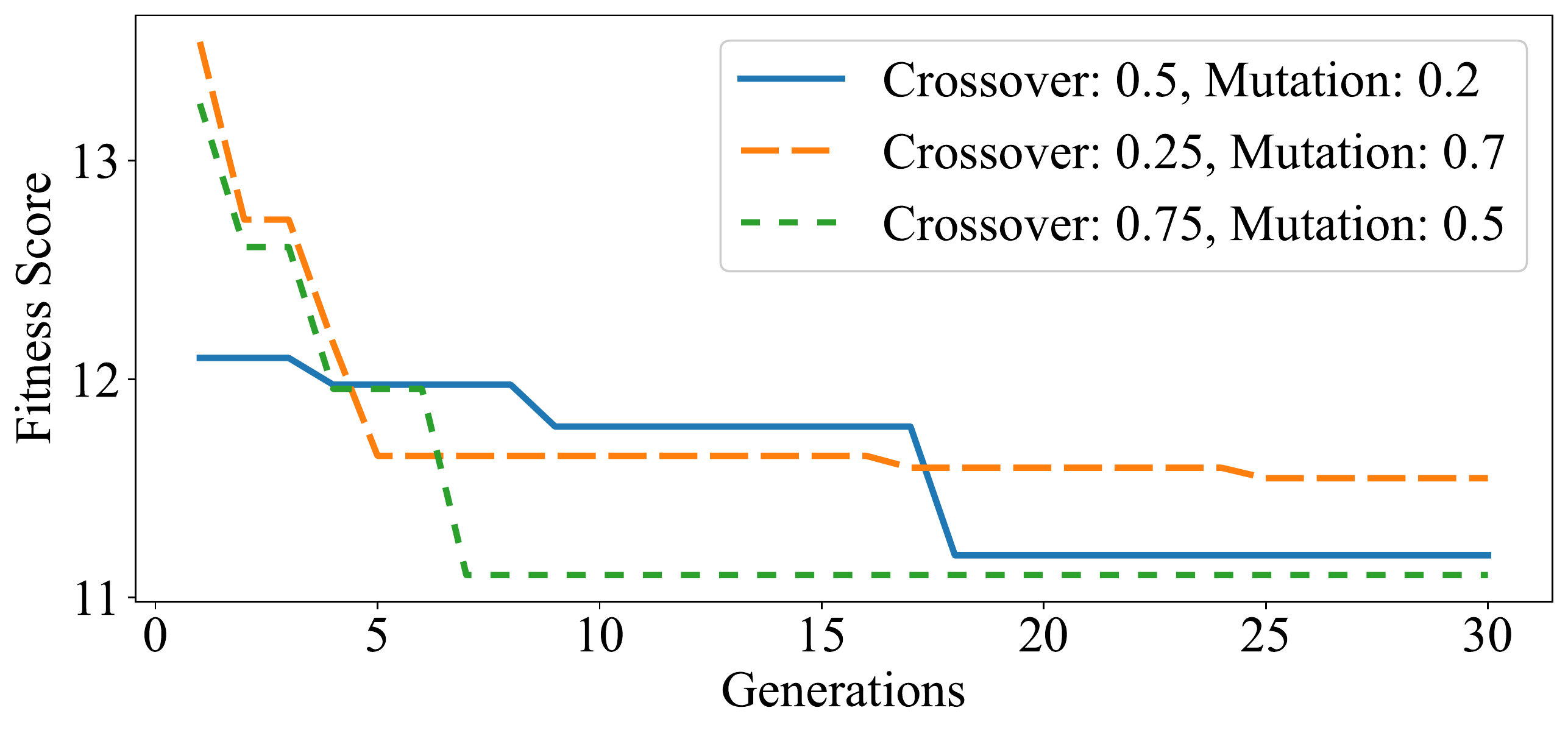}
    \caption{Performance of the best individual for different crossover and mutation probabilities in a subset of the search space.}
    \label{fig:ga_sens}
\end{figure}

In this section, we present the results of our experiments, which illustrate the added value of the proposed approach.
Our GA implementation can be found in an online repository, along with instructions for reproducing the experiments and links to the dataset that was used.\footnote{\url{https://github.com/xarakas/spheroid-tnf-v2-emews}}
For evaluation purposes, the GA was configured to run for 30 generations, with a population number of 40 individuals. 
As a selection operator, we employed the `Tournament Selection' with tournament size 3, after empirically observing that it leads to high quality solutions.
Among the various crossover operator types, we choose the Uniform crossover that applies equal probability of inheritance to all parameter values of a selected individual.
The crossover probability was set to 0.75 and the mutation probability to 0.5.
The chosen hyperparameter configuration was obtained through experimentation in a reduced search space.
Figure~\ref{fig:ga_sens} presents the evolution of the crossover and mutation probabilities for the best individual in these small scale experiments.

To illustrate the value of the proposed GA approach, we compare it against an exhaustive Sweep search, which evaluates iteratively a grid of individuals distributed uniformly in the search space.
The grid is predetermined and one simulation for each point is executed.
No particular prioritization is given to different points or subsets of the grid.
One drawback of this method when we try to balance the trade-off between reasonable execution time and amount of points evaluated, is that we may end up with a sparse grid. This may in turn lead to good simulations being skipped and never examined.
Instead,  the GA manages to focus on solutions of high quality using significantly fewer resources.

All experiments were conducted using the Mare Nostrum 4 (MN4) HPC infrastructure provided by the Barcelona Supercomputing Centre.\footnote{\url{https://www.bsc.es/marenostrum/marenostrum}}
We used up to 8 nodes with 384 processors in total.

\subsection{Simulator Calibration}

\begin{figure}
\includegraphics[width=\linewidth]{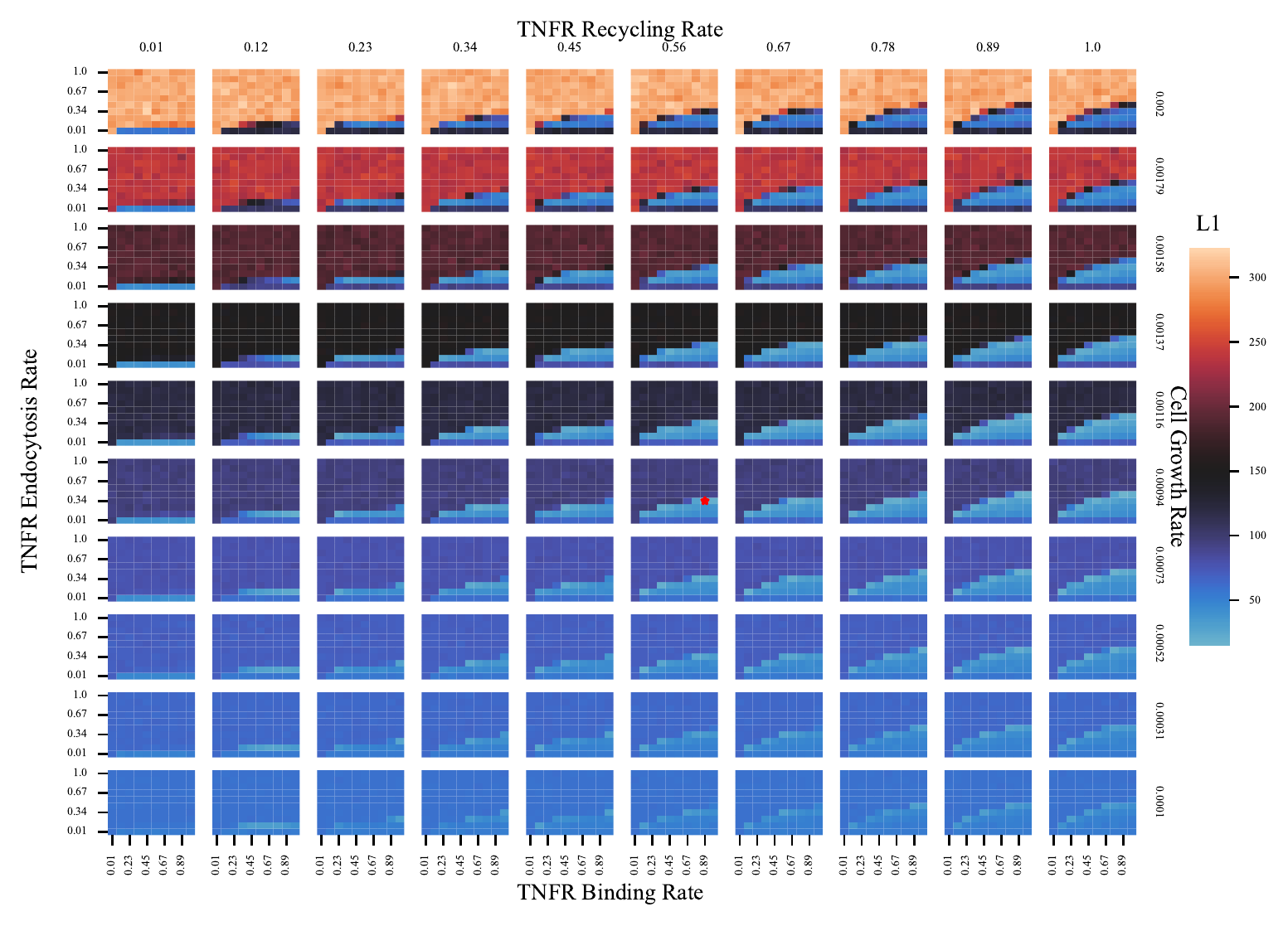}
\caption{Heatmap of all $k$ 4-tuple $L_1$ distances, as evaluated by the Sweep search. The red star denotes the configuration with the minimum score ($k_1=0.89,k_2=0.34,k_3=0.56,k_4=0.000944$, with score $L_1=14.4$). Each x-axis denotes the TNFR Binding Rate, each y-axis, the TNFR Endocytosis Rate; columns correspond to TNFR Recycling Rate values, and rows to Cell Growth Rate.}
\label{fig:3D}
\end{figure}

First, we assess the performance of Sweep and Genetic search in calibrating the simulator.
For Sweep search we create a grid of uniformly distributed points in the four dimensional space of the $k_i$ parameters and evaluate each one of them without a particular preference in the ordering of the simulations.
The $L_1$-distance scores for each parameter configuration are shown in Figure~\ref{fig:3D}, using different heatmap colors.
We observe that there are regions of the space that lead to better scores, i.e., similar results to the gold standards (light blue squares).
There are also critical regions, beyond which parameter values produce simulations that differ significantly from the gold standards (orange, red and darker blue squares).
Moreover, there seem to be ridges between `worse' and `better' regions, where locally minimum solutions are to be found. The $k$ 4-tuple with the lowest $L_1$ distance is marked with a red
star. It is worth noting that the $L_1$ distance is selected only as an illustrative example as it has more variance than the Euclidean, making the comparison more clear and accessible.

\begin{figure}

\includegraphics[width=\linewidth]{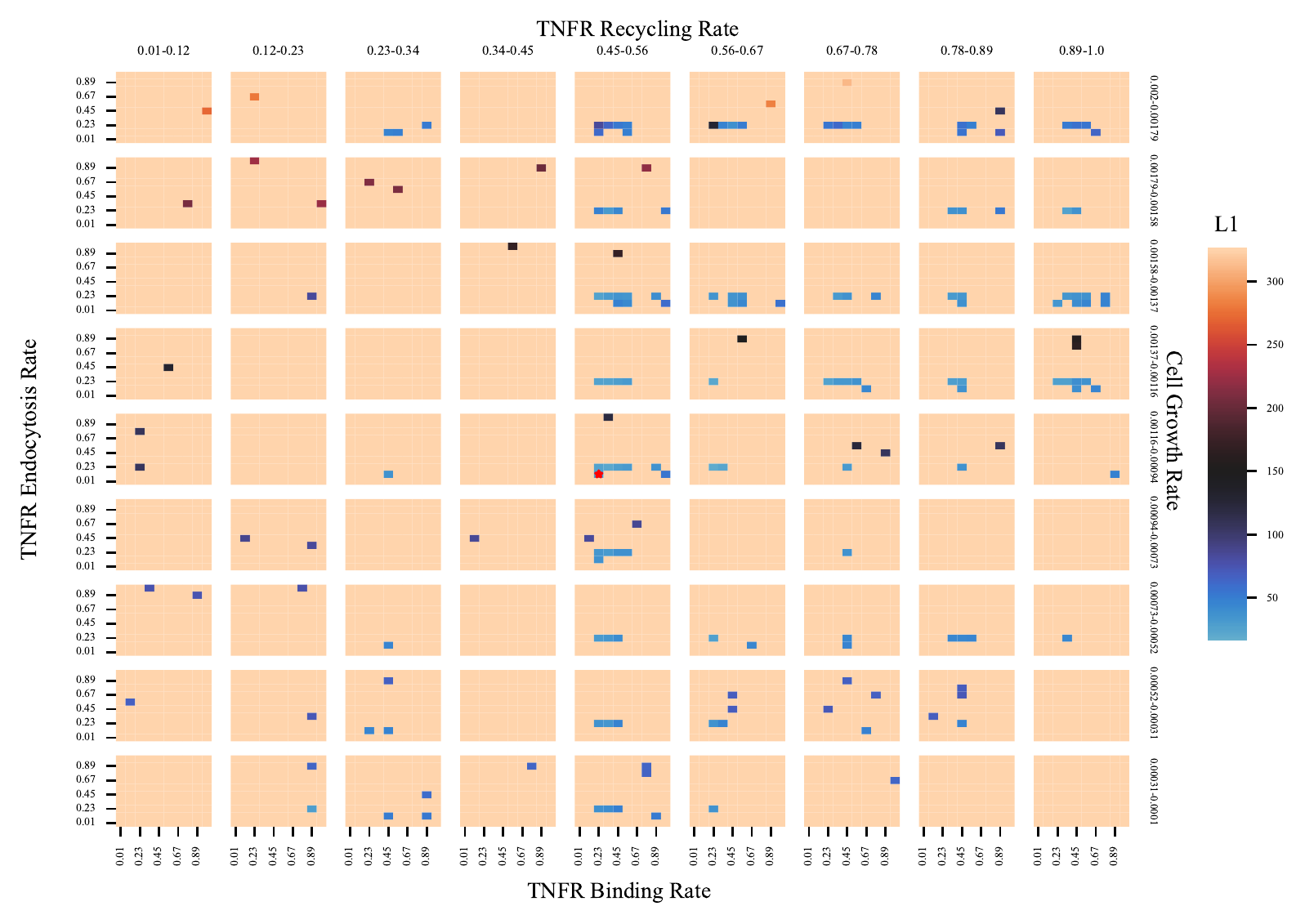}
\caption{Heatmap of all $k$ 4-tuple $L_1$ distances, as evaluated by the GA search. The red star denotes the configuration with the minimum score ($k_1=0.22,k_2=0.16,k_3=0.54,k_4=0.00097$, with score $L_1=16.28$). Each x-axis denotes the TNFR Binding Rate ranges, each y-axis, the TNFR Endocytosis Rate ranges; columns correspond to TNFR Recycling Rate ranges, and rows to Cell Growth Rate ranges. Light orange areas correspond to parts of the space that the GA did not evaluate.}
\label{fig:3DGA}
\end{figure}

In Fig.~\ref{fig:3DGA} we present the respective heatmap depicting the space visited by the GA using again the $L_1$ distance.
In order to make the results comparable to Sweep search, which examines only 10 predetermined values on each dimension, we group the GA points in the ranges set by these grid values.
The best solution is marked again with a red star, and the light orange areas correspond to value ranges that the GA did not assess.
The Figure clearly shows that  the GA evaluates a very sparse grid of points most of which are located inside the `interesting' regions of Fig.~\ref{fig:3D}.

\begin{figure}[t]
\begin{subfigure}{.48\textwidth}
  \centering
  \includegraphics[width=\linewidth]{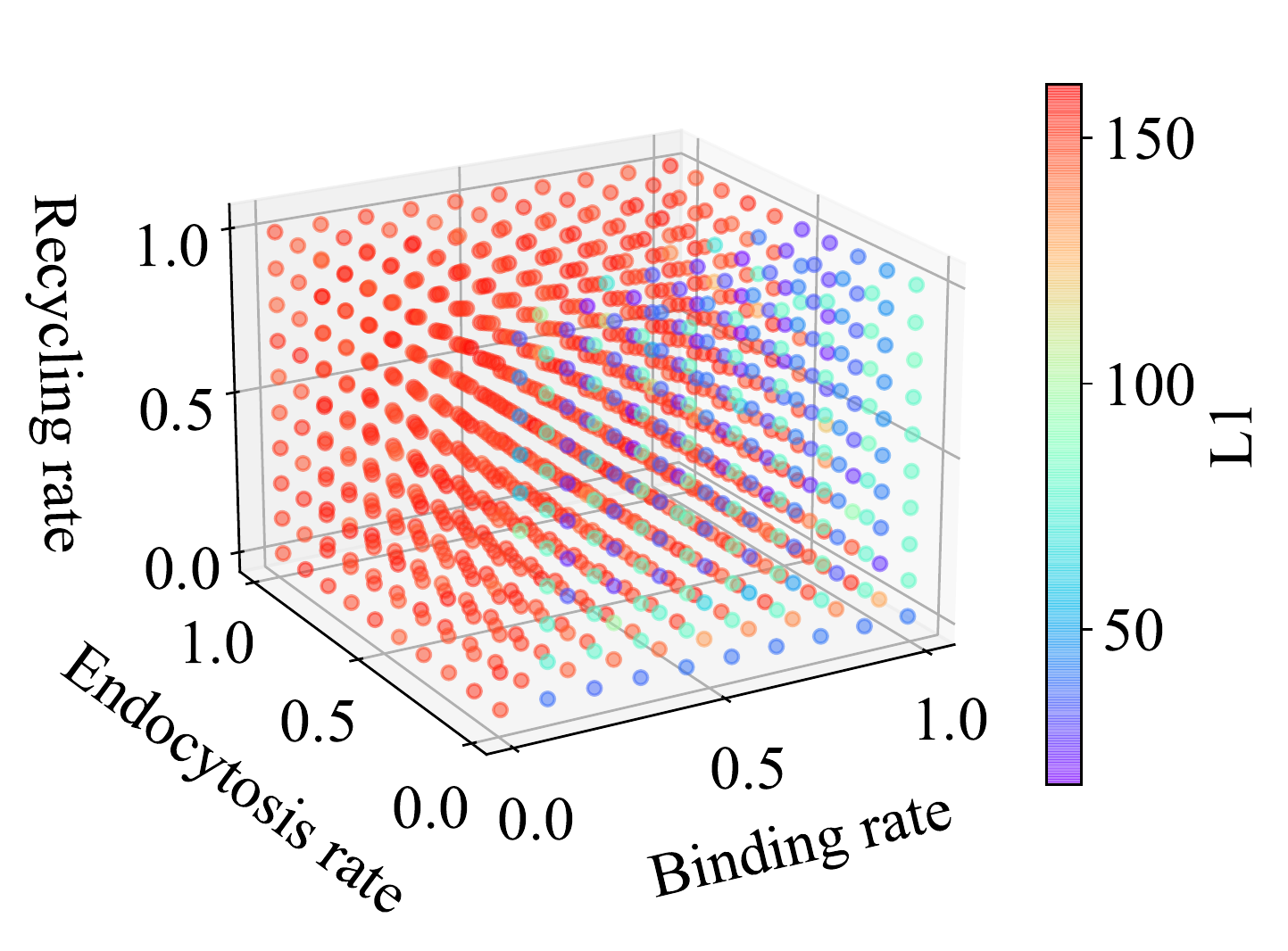}  
\end{subfigure}
\hfill
\begin{subfigure}{.48\textwidth}
  \centering
  \includegraphics[width=\linewidth]{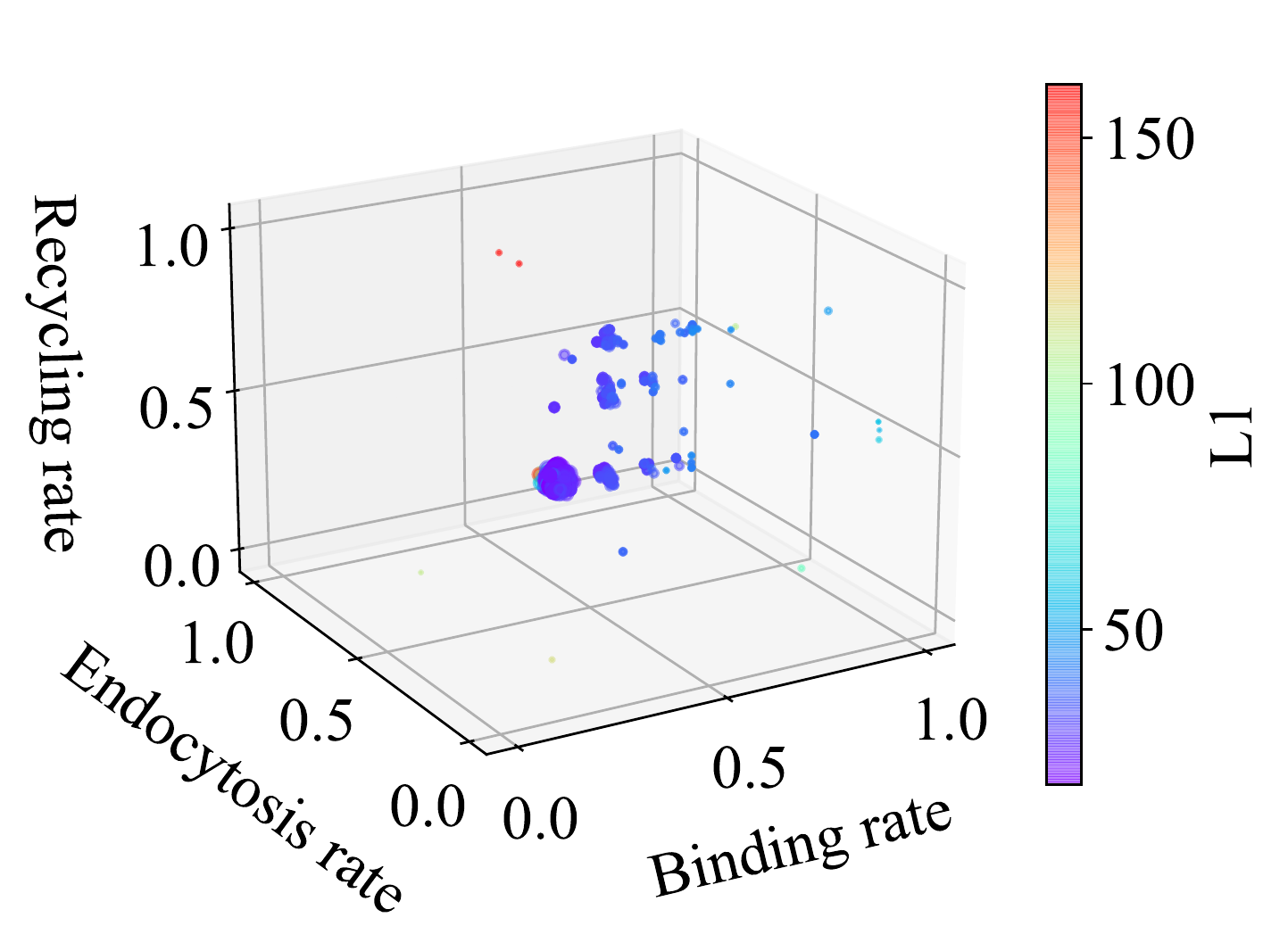}  
\end{subfigure}
\caption{Points evaluated by Sweep search (left) and GA (right), with the $k_4$ cell growth parameter lying in the range [0.0009, 0.0015]. Color denotes the $L_1$ distance from the gold standards, and point sizes in the GA case indicate temporal relationship, i.e. bigger point radius means that it is visited in later generations.}
\label{fig:3dcal}
\end{figure}

In order to gain further intuition about the behavior of the GA, we restrict the space to a 3-dimensional one in Fig.~\ref{fig:3dcal} by fixing the $k_4$ parameter to the region indicated in the fifth and sixth row of Fig.~\ref{fig:3D} and Fig.~\ref{fig:3DGA}, respectively. We show a small range instead of a particular $k_4$ value, because the GA did not visit any individual with an exact value of $k_4=0.000944$; however it did visit other individuals that have a $k_4$ that is quite close to that.
In the figure we can see that the GA avoids the evaluation of low quality configurations, by utilizing the selected fitness function and the genetic operators.
Furthermore, the GA searches subspaces with local optima, and in later generations it focuses around solutions that give promising results.

The actual parameter values selected by each method are shown in Table~\ref{tab:calresultsa}, along with the corresponding distances.
Note that absolute distance values are not directly comparable, due to the different range of each function.
Looking closer at the selected parameter values, cell growth ($k_4$) seems similar for both methods and all distance types, implying that its optimal value is  bounded.
In the case of the recycling rate ($k_3$), the selected values are again similar, with the exception of DTW for both search methods.
This is due to the fact that the recycling rate is related to periodic effects, for which the DTW distance is not strict; it is designed to bring close instances, even though them having different phases.
Finally, for the remaining parameters ($k_1$ and $k_2$), the two search methods make different selections.
This happens because the particular best point found by Sweep search is located on a ridge of local minima that are close to each other.
Overall, both methods lead to acceptable results that are close to the gold standard time-series. The solutions produced by Sweep search seems fair slightly better than those of the GA.

\begin{table}
\centering
\caption{Calibration results from Sweep search and the Genetic Algorithm}
\label{tab:calresultsa}
\begin{tabular}{|c|c|c|c|c|c|c|}
\hline
\multicolumn{2}{|c|}{Method and Distance}&Binding rate ($k_1$)&Endocytosis rate ($k_2$)&Recycling rate ($k_3$)&Cell growth rate ($k_4$)&Score\\
\hline
\multirow{3}*{\rotatebox{90}{Sweep}} 
&Euclidean& 0.89 &0.34& 0.56 &0.000944&2.53\\
&DTW& 0.89 &0.45& 1 &0.000944&7.8\\
&$L_1$& 0.89 &0.34& 0.56 &0.000944&14.4\\
\hline
\multirow{3}*{\rotatebox{90}{GA}} 
&Euclidean& 0.21 & 0.16& 0.53 & 0.00084&2.89\\
&DTW& 0.33 & 0.17 & 0.34 & 0.001&9.85\\
&$L_1$& 0.22 & 0.16 & 0.54 & 0.00097 &16.28\\
\hline
\end{tabular}

\end{table}

Turning to the computational cost of the two approaches, in Table~\ref{tab:comparison} the number of configurations ($k$ 4-tuples) evaluated by each method is shown, together with the total number of simulations that were conducted and the required time for them to complete.
Sweep search is shown in one row, because in a single run we can measure all three distances simultaneously, while the GA follows a different evolution path for each distance.
From these results, it becomes clear that the GA examines fewer points than  the Sweep search.
Since we compare against two different gold standard cases, i.e. simulation results from Letort et al.\cite{letort_physiboss:_2018}  that are considered as ground truth,  the number of simulations is twice the number of configurations examined.
In general, as the GA examines fewer configurations, it also needs fewer computational resources. 

\begin{figure}
	\centering
    \includegraphics[width=0.5\linewidth]{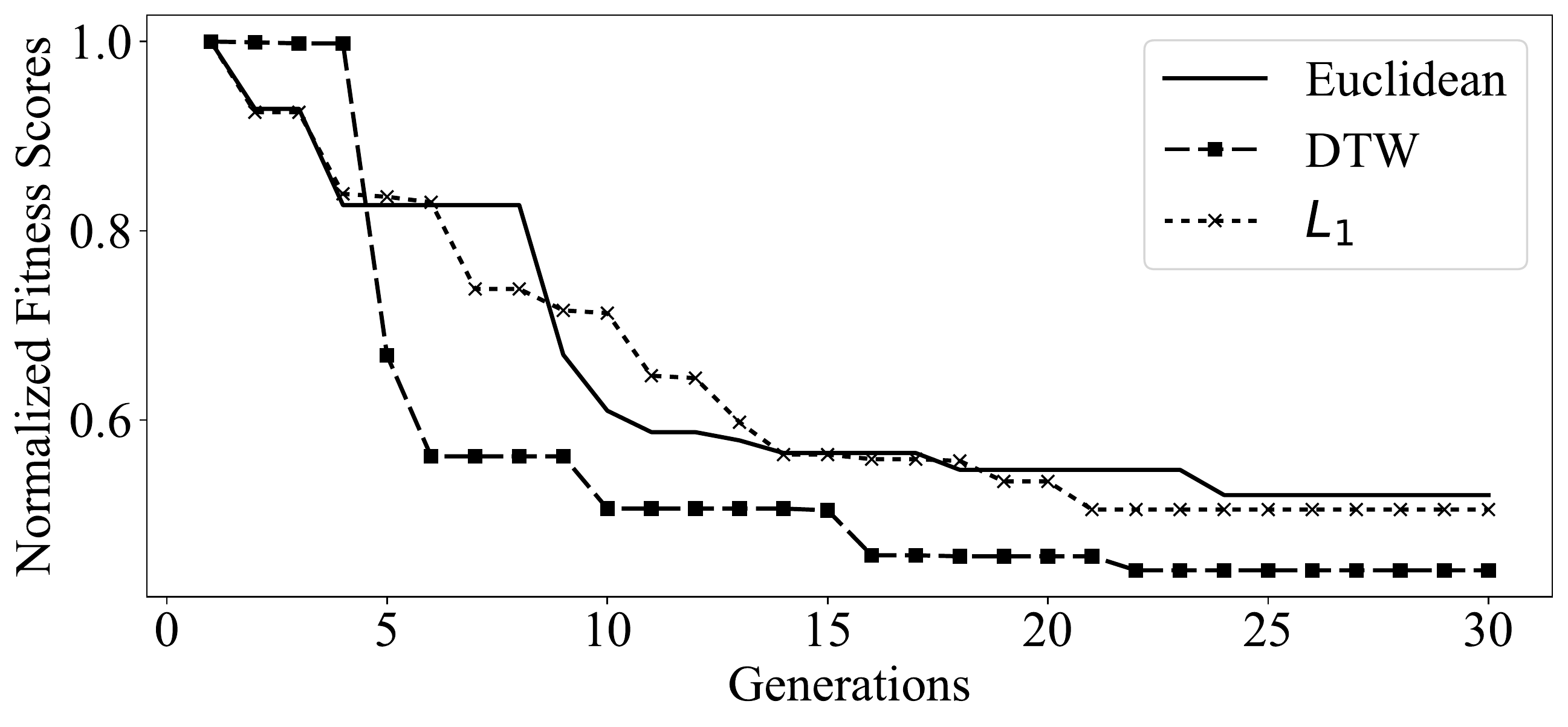}
    \caption{Genetic algorithm convergence.}
    \label{fig:ga_plot}
\end{figure}

Specifically, it performs an order of magnitude fewer simulations and with one sixth of the time spent compared to the Sweep search approach. 
Recall that the genetic search population comprises 40 individuals for 30 generations. Therefore, 1.200 individuals are assessed in each run. 
However, a proportion of these correspond to duplicate individuals, which managed to survive across generations, and do not need to be evaluated. Delving deeper into the evolution process, Figure~\ref{fig:ga_plot} depicts the normalized scores of the 
best individuals of each generation in the calibration trials. 
Despite the different magnitude of different non-normalized fitness scores for each distance type Table~\ref{tab:calresultsa}, when examining normalized scores there is a clear drop in the fitness, indicating that individuals of higher quality are examined throughout the generations.
However, using the $L_1$ distance the progress of the evolution process is smother.

\begin{table}
\centering
\caption{Comparison between Sweep search and Genetic Algorithm for the calibration experiments that used 384 CPUs each.}
\label{tab:comparison}
\begin{tabular}{|c|c|c|c|c|}
\hline
\multicolumn{2}{|c|}{Method}&$k$ 4-tuples&Simulations&Minutes\\
\hline
Sweep
&All distances& 10000 &20000& 5268\\
\hline 
\multirow{3}*{\rotatebox{90}{GA}}
&Euclidean& 940 & 1880 &  877\\
&DTW& 940 & 1880 & 881\\
&$L_1$& 871 & 1742 &  876\\
\hline
\end{tabular}

\end{table}

\begin{figure}[t]
\begin{subfigure}{.48\textwidth}
  \centering
  \includegraphics[width=\linewidth]{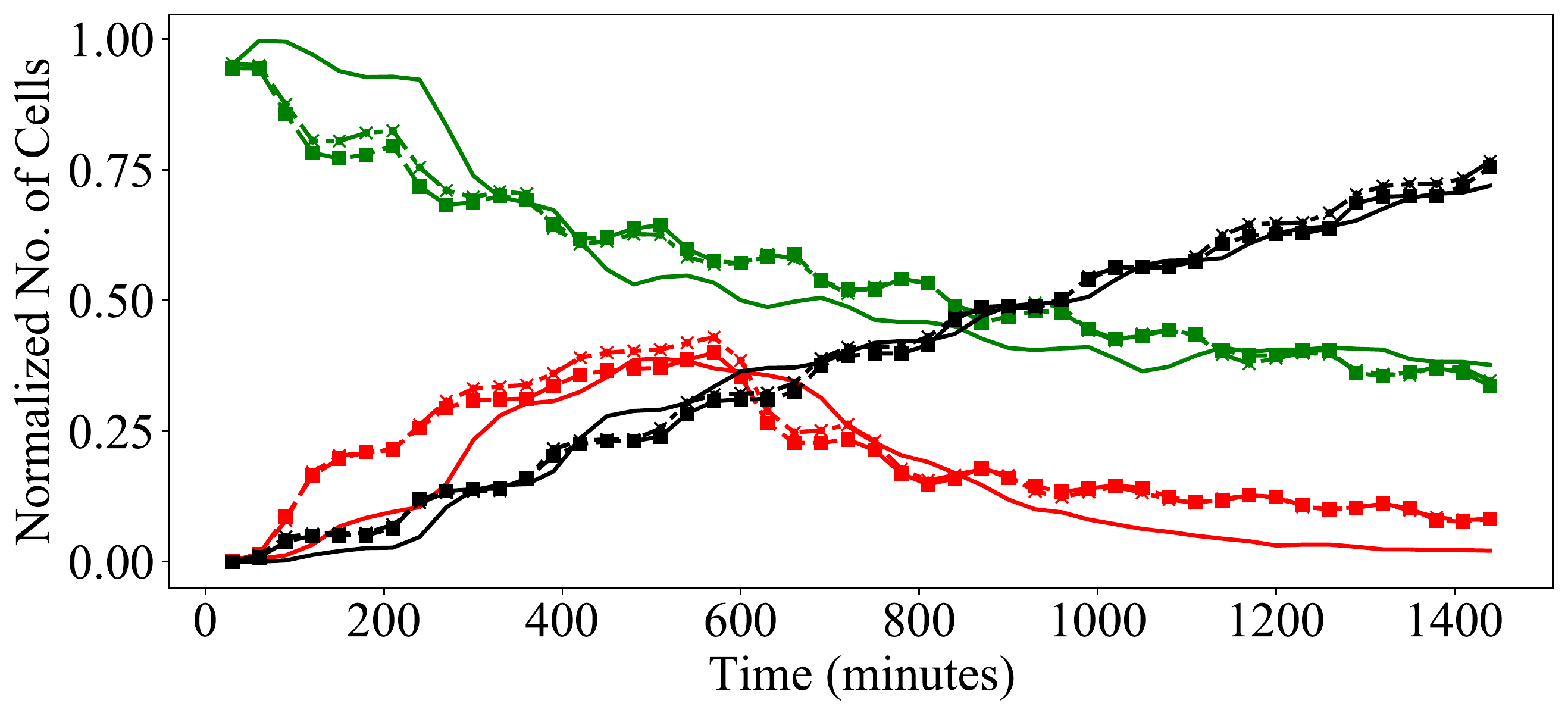}  
  \caption{Sweep search 150' TNF pulse}
\end{subfigure}
\hfill
\begin{subfigure}{.48\textwidth}
  \centering
  \includegraphics[width=\linewidth]{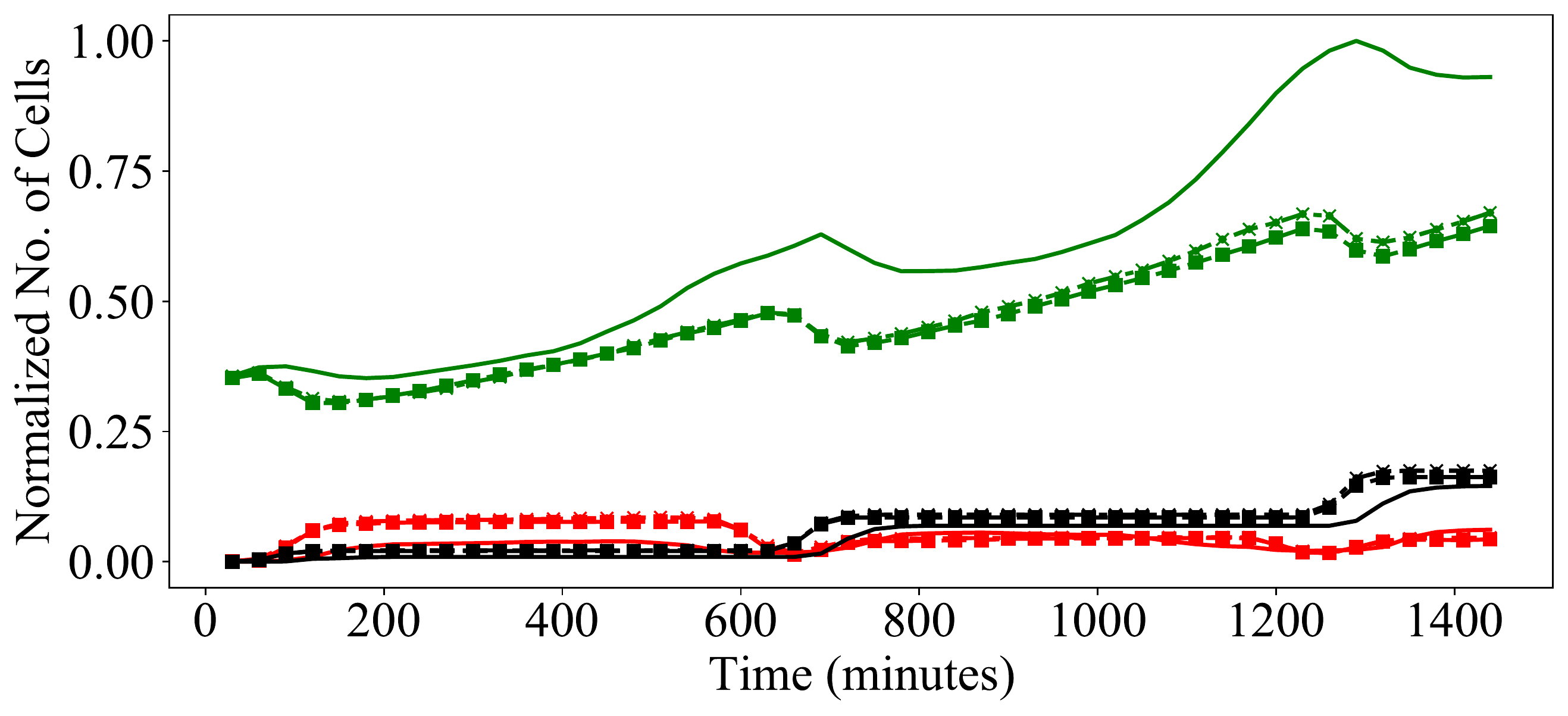}  
   \caption{Sweep search 600' TNF pulse}
\end{subfigure}
\vskip\baselineskip
\begin{subfigure}{.48\textwidth}
  \centering
  \includegraphics[width=\linewidth]{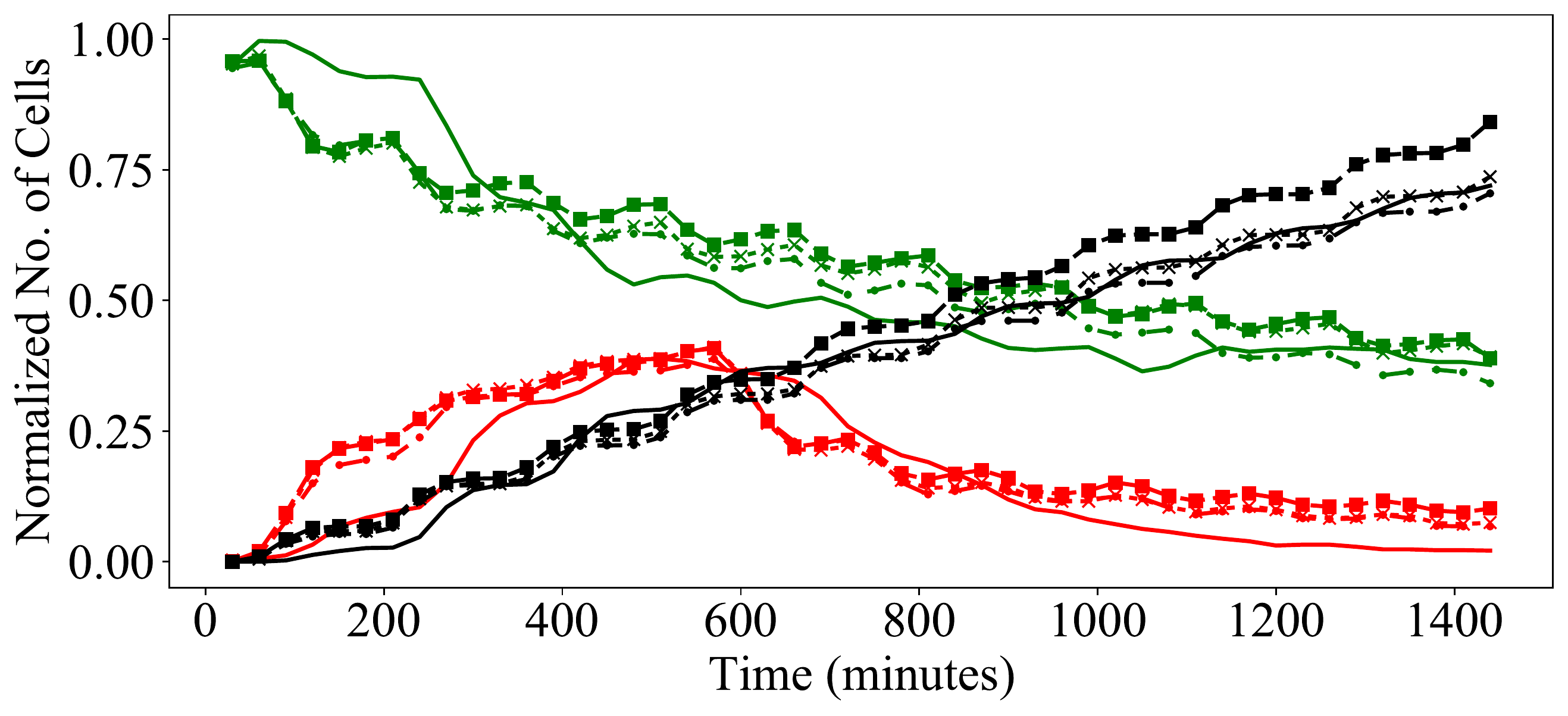}  
    \caption{Genetic Algorithm 150' TNF pulse}
\end{subfigure}
\hfill
\begin{subfigure}{.48\textwidth}
  \centering
  \includegraphics[width=\linewidth]{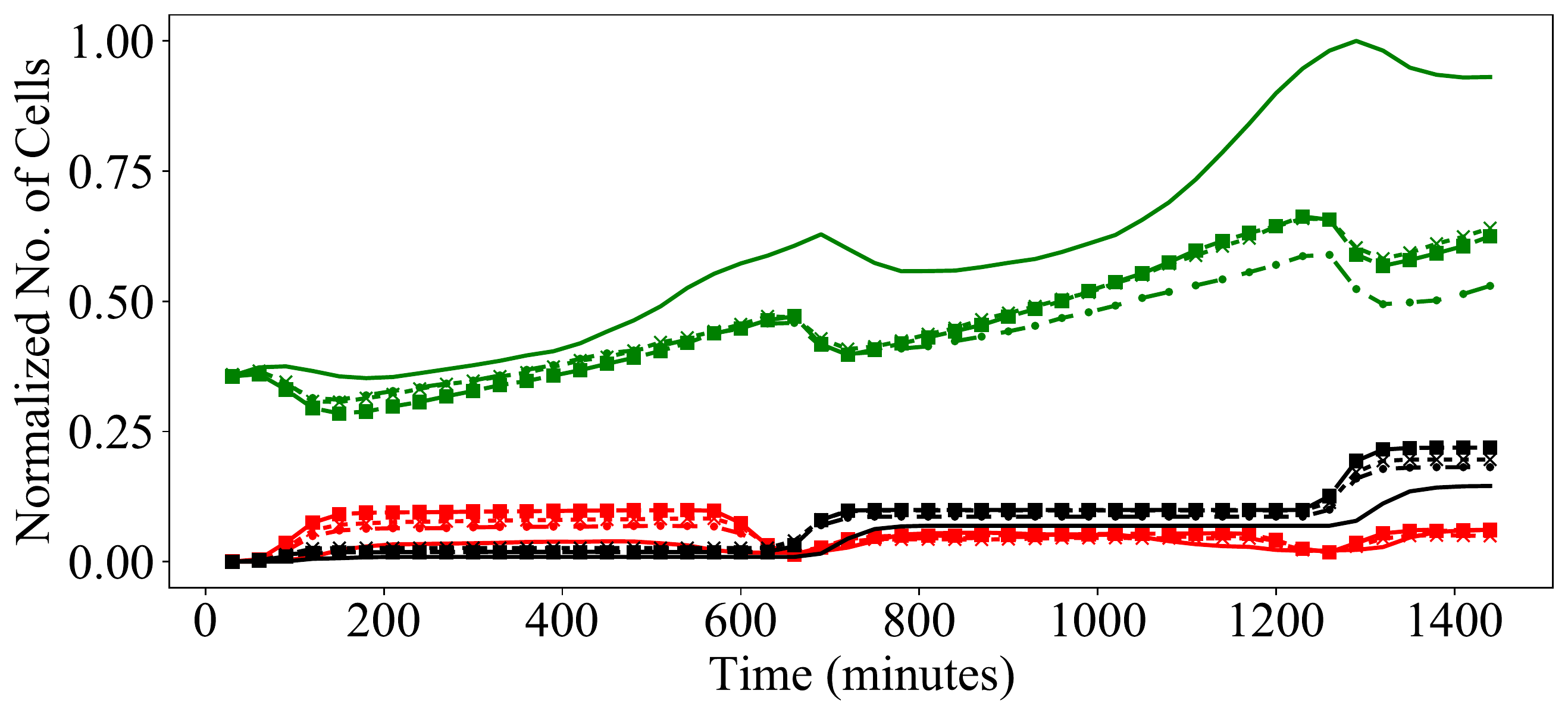}  
      \caption{Genetic Algorithm 600' TNF pulse}
\end{subfigure}
\caption{Time-series produced by the fittest $\{\textit{TNFR Binding rate}/k_1,\textit{TNFR Endocytosis rate}/k_2,$ $\textit{TNFR Recycling rate}/k_3$ $\textit{Cell growth rates}/k_4\}$ value combinations. Top: Sweep search; Bottom: Genetic Algorithm. Left: 150' TNF pulse (Gold Standard \#1 case); right: 600' TNF pulse (Gold Standard \#2 case). Green curves show the number of alive cells, red the number of apoptotic cells, and black the number of necrotic cells. Solid lines are the gold standards. Dot-marked curves correspond to the solution provided by the Euclidean distance, squared-marked curves correspond to those computed with the use of the DTW distance, and `x'-marked curves computed with the use of the $L_1$ distance.}
\label{fig:TS}
\end{figure}

As a final step in our analysis, we inspect visually the actual time-series produced by the simulator using the values selected by the search methods, against the two gold standards.
The results are presented in Figure~\ref{fig:TS}.
Looking at the gold-standard, solid curves, for the TNF=150 injection strategy (left figures), we can see that the number of alive cells fluctuating every 150 minutes, which is the administration frequency, but nevertheless achieving a decreasing trend. 
Respectively, necrotic cell numbers are increasing as time passes.
The apoptotic cells increase in the first half of the simulation to later follow the drop of the alive cell count.
In the second injection strategy, TNF=600 (right figures), the drug administration does not have the desired effect, since the number of alive tumor cells grows larger than at the beginning of the simulation.

Regarding the behavior of the solutions selected by the search methods, we first examine the case of Sweep search (top figures).
For the TNF=150 gold standard (left figure), the chosen configuration behaves similarly to the golden one, independent of the distance function that is used.
The same holds for the TNF=600 case (right figure), especially in the first half of the simulation. The number of alive cells seems to be underestimated in the second half of the simulation.
Overall, the simulator seems to be calibrated successfully by the search methods.

\subsection{Drug policy exploration}
Having selected the best PhysiBoSSv2.0 configuration for the four rate values, we now focus our exploration on drug treatment characteristics.
In particular, the variables we wish to tune are the frequency of TNF injections, the duration and the concentration of TNF dosages given to patients, i.e. each individual is represented as $\{\textit{TNF Administration Frequency}/k_5,\textit{TNF Duration}/k_6,$ $\textit{TNF concentration}/k_7\}$. 
Since in this stage there is no gold standard case to match against, the fitness function counts the number of `Alive' cells at the final time step of each simulation---which in our case is after 1440 minutes of treatment.

\begin{figure}
\begin{subfigure}{.48\textwidth}
  \centering
  \includegraphics[width=\linewidth]{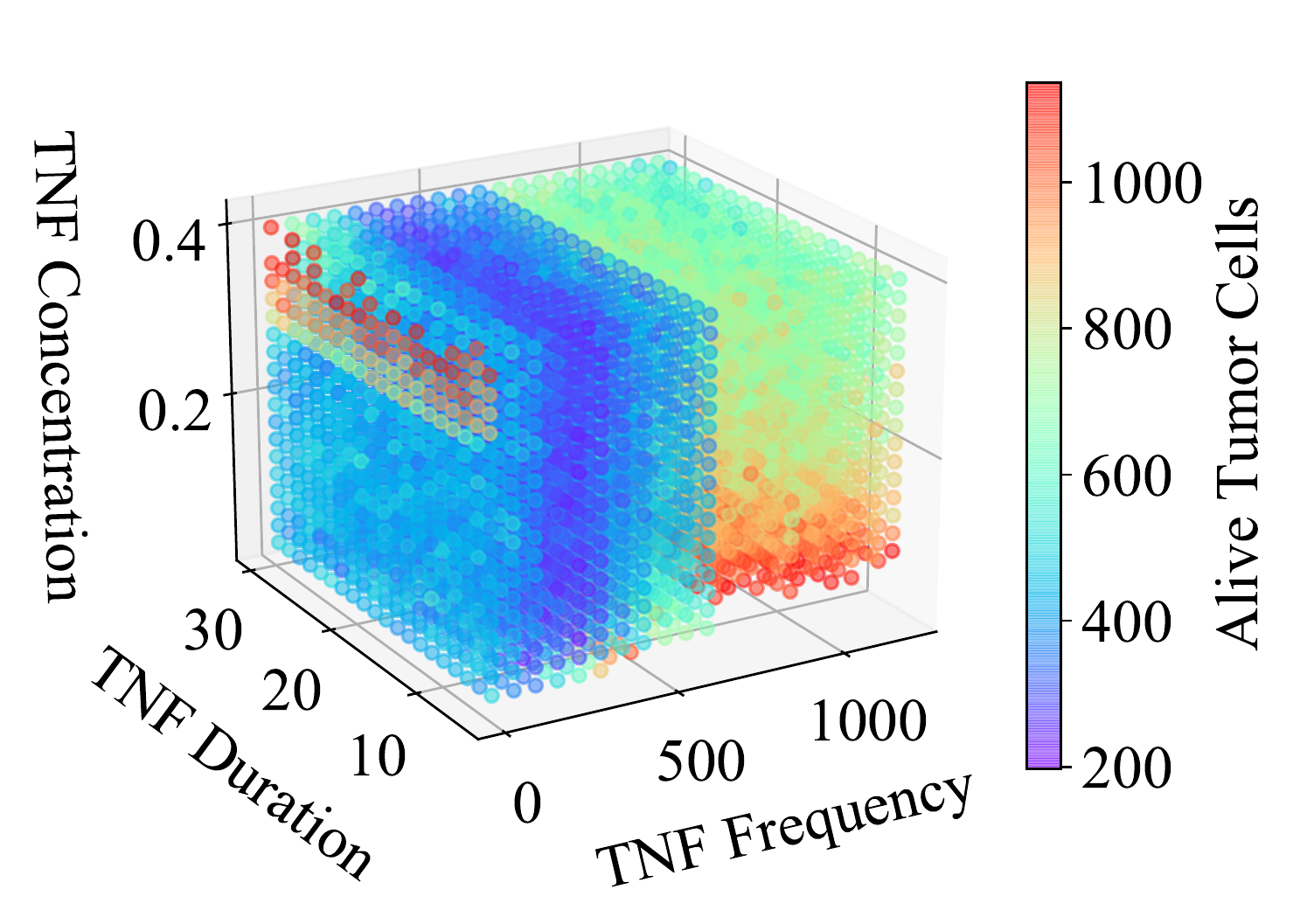}  
  \caption{Sweep search}
\end{subfigure}
\hfill
\begin{subfigure}{.48\textwidth}
  \centering
  \includegraphics[width=\linewidth]{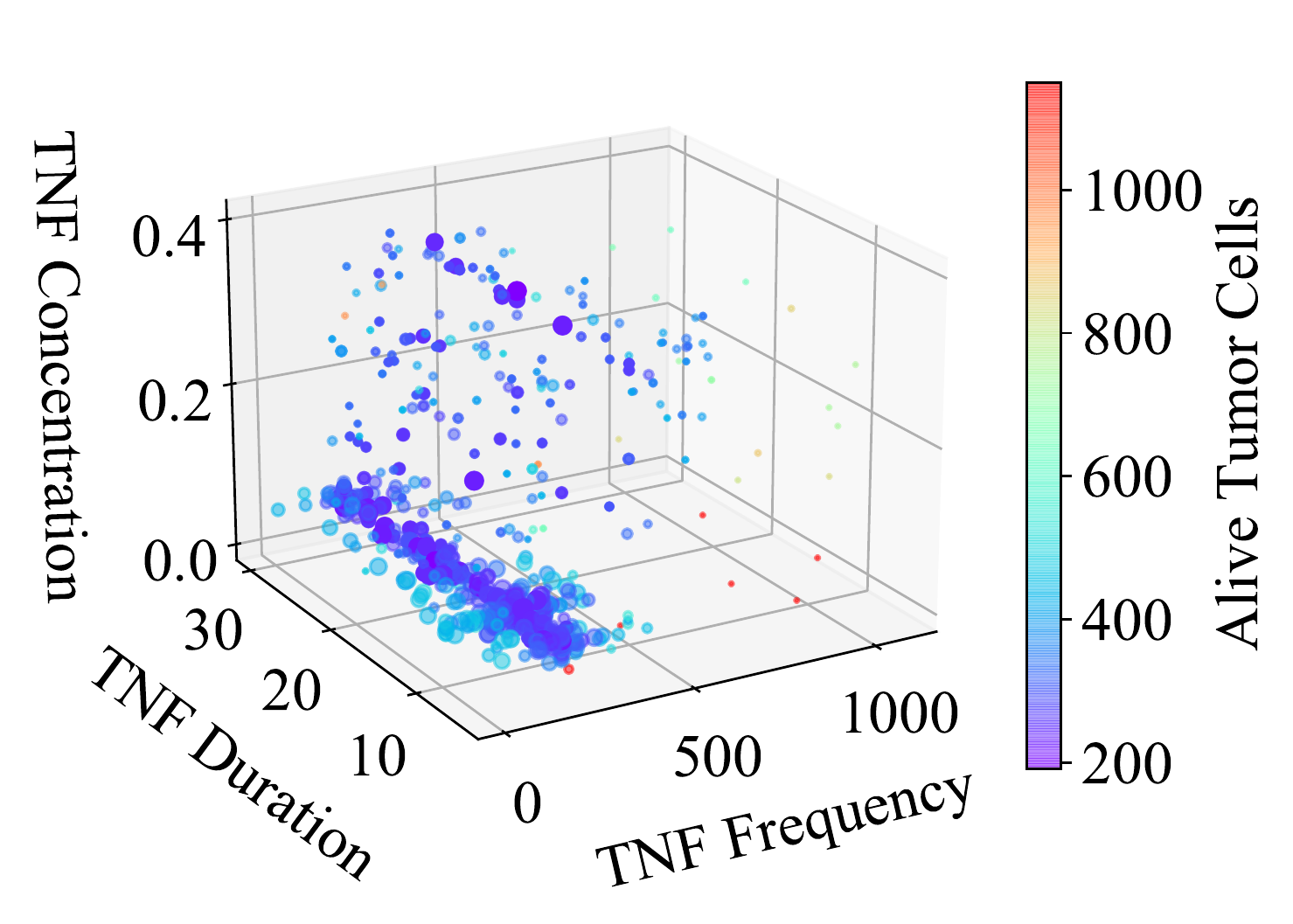} 
  \caption{Genetic Agorithm}
\end{subfigure}
\caption{$k$-triple evaluations for drug policy exploration, Sweep search on the left, and GA on the right. Color denotes the number of alive tumor cells at the end of the simulation.}
\label{fig:DD3ds}
\end{figure}

Similar to the previous set of experiments, we first perform a Sweep search of the space to identify interesting regions.
Also, to further simplify the presentation, we distinguish solution candidates to viable and unviable ones.
The latter are those that end up with more alive tumor cells at the end of the simulation than at the beginning, meaning that the particular drug treatment is not successful. These solutions are omitted from the figures. 
On the other hand, the viable ones are those that manage to reduce the tumor cell count.

The results of Sweep search are shown on the left of Figure~\ref{fig:DD3ds}, marking with a different color each individual according to its fitness score.
As we can see, there is an interesting region that contains parameter values, which lead to promising drug treatment configurations (dark blue and purple points).
The GA on the other hand, as shown on the right of  Figure~\ref{fig:DD3ds} converges to points that lie within the interesting region marked by the Sweep search.
Again, for the GA case, the size of each point increases proportionally to the generation that it is examined, meaning that larger points are visited later  in the evolution process.
Thus, the GA is able to distinguish among good and bad solutions and expand the search around potentially interesting regions of the search space.
The actual parameter values produced by each of the two approaches is shown in Table~\ref{tab:calresults}, with the GA pointing to a solution that leads to fewer tumor cells (191) than the Sweep search (197). 

\begin{table}
\centering
\caption{Drug policy exploration results of the Sweep search and the Genetic Algorithm.}
\label{tab:calresults}
\begin{tabular}{|l|c|c|c|c|}
\hline
Method &TNF Administration Frequency ($k_5$)&TNF Duration ($k_6$)&TNF Concentration ($k_7$)&No. of Alive cells\\
\hline
Sweep& 260.52 &20.78& 0.067 &197\\
GA& 192 & 19&0.0457 &191\\
\hline
\end{tabular}

\end{table}

Table~\ref{tab:comparison2} presents the number of configurations examined, which is the same as the number of simulations performed in this case, and the computation time for each case.
The GA manages to arrive at good solutions of lower `Alive' cell counts, by examining one order of magnitude fewer candidate solutions than Sweep search, in less than half of the time.

\begin{table}
\centering
\caption{Comparison of the Sweep search and the GA for the drug policy exploration experiment. Each, used 384 CPUs.}
\label{tab:comparison2}
\begin{tabular}{|l|c|c|c|}
\hline
Method&$k$ triples&Simulations&Minutes\\
\hline
Sweep& 8000 &8000&2809\\
GA& 632 & 632& 1120\\
\hline
\end{tabular}

\end{table}

\section{Conclusions \& Further Work}
Multi-scale simulations have been proven quite valuable in various application areas and the domain of drug discovery is no exception.
In this work, we incorporated PhysiBoSSv2.0 into EMEWS workflows to allow the parallel execution of large-scale cancer cell growth experiments.
Our goal was first to configure the new version of the simulator properly in order to produce useful results and then to use it for discovering effective drug treatment policies that are able to attack cancer cells and contain their growth.
Both for the model calibration and for the drug discovery stage, we proposed the use of a Genetic Algorithm that can search the space efficiently utilizing a fitness function to characterize the suitability of candidate solutions.
We assessed different types of fitness functions and analyzed their ability to lead to good solutions.
Our experimental evaluation, performed on a high performance computing infrastructure, confirm the validity of the proposed approach, highlighting the difference between informed and uninformed search strategies. 

Moving forward, we plan to assess additional AI methods and compare their performance against the GA while also expanding our search to other simulator parameters.
Moreover, we are developing an approach for the early termination of non-promising simulations, which seem to deviate from the objectives of the experiment.
This could be achieved, for example, with the incorporation of early time-series classification techniques.\cite{wayeb,ets} 
Furthermore, our approach is  incorporated in the HORIZON-2020 INFORE project in order to serve as a sub-module in a larger system designed for extreme-scale analytics.

\section*{Acknowledgements}
This work has received funding from the EU Horizon 2020 RIA program INFORE under grant agreement No 825070. 
\bibliographystyle{splncs04}
\bibliography{bibliography/bsc,bibliography/ncsr}
\end{document}